\documentclass[article]{jss}

\usepackage{thumbpdf,lmodern}
\usepackage{graphicx}
\usepackage{subfig}
\usepackage{mathtools, nccmath}
\usepackage{array}
\newcolumntype{P}[1]{>{\centering\arraybackslash}p{#1}}

\usepackage{framed}
\usepackage{amsmath}
\usepackage{amssymb}

\newcommand{\fct}[1]{\code{#1()}}

\author{Jane Pan\\University of California, Los Angeles \And 
Sudipto Banerjee\\University of California, Los Angeles}
\Plainauthor{Jane Pan, Sudipto Banerjee}

\title{bayesassurance: An R Package for 
Calculating Sample Size and Bayesian Assurance}
\Plaintitle{bayesassurance: An R Package for 
Calculating Sample Size and Bayesian Assurance}
\Shorttitle{Bayesian Assurance in \proglang{R}}

\Abstract{
We present a \pkg{bayesassurance} 
\proglang{R} package that computes
the Bayesian assurance under various 
settings characterized by different assumptions and 
objectives.  The package offers a constructive set 
of simulation-based functions suitable for addressing 
a wide range of clinical trial study design problems.  
We provide a detailed description of the underlying 
framework embedded within each of the power and assurance 
functions and demonstrate their usage through a series
of worked-out examples. Through these examples, we
hope to corroborate the advantages that come 
with using a two-stage generalized
structure. We also illustrate scenarios where the
Bayesian assurance and frequentist power overlap,  
allowing the user to address both Bayesian and
classical inference problems provided that the parameters
are properly defined. 
All assurance-related functions included in this 
\proglang{R} package rely on a two-stage 
Bayesian method that assigns two distinct 
priors to evaluate the unconditional 
probability of observing a positive outcome,  
which in turn addresses subtle limitations that 
take place when using the standard
single-prior approach. 
}


\Address{
Jane Pan\\
Department of Biostatistics\\
University of California, Los Angeles\\
650 Charles E. Young Drive South, Los Angeles, CA 90095\\
E-mail: \email{jpan1@ucla.edu}
}

\Address{
Sudipto Banerjee\\
Department of Biostatistics\\
Faculty and Chair of Biostatistics\\
University of California, Los Angeles\\
650 Charles E. Young Drive South, Los Angeles, CA 90095\\
E-mail: \email{sudipto@ucla.edu}
}

\begin{document}
\section[Introduction: Bayesian Assurance and Sample Size Determination]{Introduction} \label{sec:intro}

Power and sample size analysis is
an essential component for developing
well-structured study designs. In the classical setting, identifying 
an appropriate sample size to achieve
a desired statistical power offers valuable insight when 
making decisions. Power curves serve
as a useful tool in quantifying the degree of assurance
held towards meeting a study's analysis objective
across a range of sample sizes. 
Formulating sample size determination as a 
decision problem so that power is an increasing 
function of sample size, offers the investigator a visual aid in
helping deduce the minimum sample size needed to 
achieve a desired power. 

The overarching idea of Bayesian assurance 
relies on measuring the probability
of meeting a desired objective. 
The Bayesian setting has allocated a decent amount 
of attention towards sample size determination, 
accumulating a diverse collection
of methods stemming from different perspectives
over the last few decades. Within this vast collection
of methodologies contains a class of approaches that 
determine sample size based upon some criteria 
of analysis or model performance
(\cite{rahme}; \cite{wang}; \cite{ohagan}). Other 
proposed solutions of the sample size problem 
take on a more tailored approach and introduce 
frameworks that prioritize conditions specific to the 
problem at hand, e.g. the use of Bayesian average 
errors to simultaneously control for Type I
and Type II errors \citep{reyes}, the use of
posterior credible interval lengths to gauge
sample size estimates \citep{joseph}, and the
use of survival regression models to target
Bayesian meta-experimental designs \citep{reyes}.

\begin{table}[t!]
\centering
\begin{tabular}{lllp{6.2cm}}
\hline
Function  & Type & & Description \\ \hline
\fct{pwr\_freq} & closed-form solution & &
Returns the statistical power
of the specified hypothesis test (either
one or two-sided).  \\
\fct{assurance\_nd\_na} & closed-form solution & & Returns 
the exact Bayesian assurance for 
attaining a specified alternative.\\
\fct{bayes\_sim} & simulation & & Approximates the Bayesian assurance of attaining a specified
condition for a balanced study design through 
Monte Carlo sampling.\\
\fct{bayes\_sim\_unbalanced} & simulation & & 
Approximates the Bayesian assurance of attaining 
a specified condition for an unbalanced study 
design through Monte Carlo sampling.\\
\fct{bayes\_sim\_unknownvar} & simulation & & 
Same as \fct{bayes\_sim} but assumes the
variance is unknown.\\
\fct{bayes\_adcock} & simulation & & 
Determines the assurance of observing that the
absolute difference between the true underlying
population parameter and the sample estimate falls within a margin of error no greater than a fixed precision level, $d$.\\
\fct{bayes\_sim\_betabin} & simulation & &
Returns the Bayesian assurance corresponding to a hypothesis test for difference in two independent proportions.\\
\fct{pwr\_curve} & visual tool & & Constructs a plot
with the power and assurance curves overlayed on top
of each other for comparison. \\
\fct{gen\_Xn} & design tool & &
Constructs design matrix using given sample size(s).
Used for power and sample size analysis in the Bayesian setting. \\
\fct{gen\_Xn\_longitudinal} & design tool & &
Constructs design matrix using inputs that correspond
to a balanced longitudinal study design.\\
 \\ \hline
\end{tabular}
\caption{\label{tab:overview} Overview of the
functions available for use within the package.}
\end{table}

There are currently several R packages 
pertaining to Bayesian sample size calculations, each
targeting a different set of study design approaches. 
The \pkg{SampleSizeMeans} package produced by 
Lawrence Joseph and Patrick Belisle contains
a series of functions used for determining appropriate
sample sizes based on various Bayesian criteria 
for estimating means or differences between means
within a normal setting \citep{samplesizemeans}. Criteria
considered include the Average Length Criterion, 
the Average coverage criterion, and the 
Modified Worst Outcome Criterion 
\citep{joseph1995, joseph1997}. 
A supplementary package, \pkg{SampleSizeProportions}, addresses study designs for estimation of binomial
proportions using the same set of criteria listed
\citep{samplesizeprop}. 
Our package, also
addresses the application of criteria based within 
the normal and binomial setting, places strong 
emphasis on a
two-stage framework primarily within the
context of conjugate Bayesian linear 
regression models, where we consider the situation
with known and unknown variances.  Our R package also
offers the flexibility of considering unequal sample sizes
for two samples as well as longitudinal study designs. 


The \pkg{bayesassurance} package 
contains a collection of
functions that can be divided into three categories
based on design and usage. These include closed-form solutions, simulation-based solutions, and 
visualization and/or design purposes. All available functions are categorized and 
briefly described in Table~\ref{tab:overview}.
The manuscript is organized as follows.
Section~\ref{sec:closedformassurance} introduces
one of the more basic features included 
in the package that is linked to
the fundamental closed-form solution of assurance. 
This section also introduces the notion of overlapping
behaviors exhibited between the Bayesian 
and frequentist settings, a concept that is 
frequently brought up and discussed
throughout the paper.
Sections~\ref{sec:simform} and~\ref{sec:othermethods}
explore simulation-based assurance methods, 
outlining the statistical framework associated with 
each method followed by
examples worked out in \proglang{R} 
that users could replicate. 
Section~\ref{sec: goalfunc} discusses a related topic 
similar to the assurance that determines sample size in 
relation to the rate of correct classification. Here we discuss 
how a Bayesian goal function is formulated and applied
in the context of sample size determination. 
Section~\ref{sec:tools} offers some useful graphical 
features and design matrix generators embedded within
the assurance-based functions. 
In the following sections, we provide a detailed 
overview for each of the functions available grouped 
by category followed with worked out examples in R.

\section{Closed-form Solution of Assurance}\label{sec:closedformassurance}

The Bayesian assurance evaluates the tenability of 
attaining a specified outcome through the implementation
of prior and posterior distributions. 
The \fct{assurance\_nd\_na} function computes the
exact assurance using a closed-form
solution.  

The setup of the assurance framework is comparable 
to that of the classical power setting but avoids the 
hypothesis testing structure and refrains from 
the use of absolute phrases such as ``rejecting'' or 
``failing to reject the null hypothesis''. 
Instead, the Bayesian paradigm defines an objective 
that users wish to measure the probability of attaining. 
Suppose we seek to evaluate the tenability 
of $\theta > \theta_0$ given data from a Gaussian 
population with mean $\theta$ and known variance
$\sigma^2$. We assign two sets of priors for $\theta$, 
one at the $\textit{design stage}$ and the other at the 
$\textit{analysis stage}$. These two stages are the 
primary components that make up the skeleton of our 
generalized solution in the Bayesian setting and will be
revisited in later sections. The analysis objective specifies 
the condition that needs to be satisfied. It defines a positive
outcome, which serves as an overarching criteria
that characterizes the study. 
In this setting, the analysis objective is to observe 
$P(\theta > \theta_0| \bar{y}) > 1-\alpha$. 
The design objective seeks a sample size that is 
needed to ensure that the analysis objective is met 
$100\delta\%$ of the time, where $\delta$ denotes 
the assurance.

Let $\theta \sim N\big(\theta_1, \frac{\sigma^2}{n_a}\big)$ 
be our analysis stage prior and $\theta \sim N\big(\theta_1,
 \frac{\sigma^2}{n_d}\big)$ be our design stage prior, 
where $n_a$ and $n_d$ are precision parameters
that quantify the degree of belief towards 
parameter $\theta$ and the population from which 
we are drawing samples from to evaluate $\theta$.
Then given the likelihood $\bar{y} \sim N\big(\theta, 
\frac{\sigma^2}{n}\big)$, we can obtain the posterior 
distribution of $\theta$ by multiplying the 
analysis prior and likelihood:
\begin{equation}\label{eq: simple_posterior}
\nonumber
N\left(\theta {\left | \theta_1, \frac{\sigma^2}{n_a} \right.}\right) \times N\left(\bar{y} {\left | \theta, \frac{\sigma^2}{n} \right.}\right)
\propto N\left(\theta {\left | \frac{n_a}{n + n_a}\theta_1 + \frac{n}{n + n_a}\bar{y}, \frac{\sigma^2}{n + n_a}\right.}\right)\;.
\end{equation}

This posterior distribution gives us $P(\theta > \theta_0 | \bar{y})$
and the assurance is then defined as
\begin{equation}
\nonumber
\delta = P_{\bar{y}}\left\{\bar{y}: P(\theta > \theta_0 | \bar{y}) > 1 - \alpha\right\}.
\end{equation}
The assurance expression can be expanded out further by 
using the marginal distribution of $\bar{y}$, which is obtained by
\[
\int{N\left(\theta {\left|\theta_1, \frac{\sigma^2}{n_d}\right.}\right) \times N\left(\bar{y} {\left | \theta, \frac{\sigma^2}{n} \right.}\right) d\theta} 
= N\left(\bar{y} {\left |\theta_1, \Big(\frac{1}{n} + \frac{1}{n_d}\Big) \sigma^2 \right.}\right) .
\]
Since the assurance definition is conditioned on $\bar{y}$, we use this
to standardize the assurance expression to obtain the following
closed-form solution: 
\begin{equation}
\label{eq:assurance}
\delta(\Delta, n) = \Phi\Bigg(\sqrt{\frac{nn_d}{n+n_d}}
\Bigg[\frac{n+n_a}{n}\frac{\Delta}{\sigma} + Z_{\alpha}\frac{n+n_a}{n}\Bigg]\Bigg).
\end{equation}

The \fct{assurance\_nd\_na} function requires the following 
specified parameters: 
\begin{enumerate}
 \item{\code{n}: sample size (either scalar or vector)}
 \item{\code{n_a}: precision parameter within the analysis
 stage that quantifies the degree of belief carried towards 
 parameter $\theta$}
 \item{\code{n_d}: precision parameter within the design stage
 that quantifies the degree of belief of the population from 
 which we are generating samples from}
 \item{\code{theta_0}: initial parameter value provided by the client}
 \item{\code{theta_1}: prior mean of $\theta$ assigned in the
analysis and design stage}
 \item{\code{sigsq}: known variance}
\item{\code{alt}: specifies alternative test case, 
where \code{alt = "greater"} tests if 
$\theta_1 > \theta_0$,\code{alt = "less"} tests if 
$\theta_1 < \theta_0$, and \code{alt = "two.sided"}
performs a two-sided test for 
$\theta_1 \neq \theta_0$.  By default, 
\code{alt = "greater"}}.
 \item{\code{alpha}: significance level}
 \end{enumerate}
Consider the following code that loads in the 
\pkg{bayesassurance} package and assigns arbitrary 
parameters to \fct{assurance\_nd\_na} prior to executing
the function. 
\begin{CodeChunk}
\begin{CodeInput}

R> library(bayesassurance)

R> n <- seq(100, 250, 10)
R> n_a <- 10
R> n_d <- 10
R> theta_0 <- 0.15
R> theta_1 <- 0.25
R> sigsq <- 0.30

R> out <- assurance_nd_na(n = n, n_a = n_a,  n_d = n_d, 
	theta_0 = theta_0, theta_1 = theta_1, sigsq = sigsq,  
	alt = "greater", alpha = 0.05)

R> head(out$assurance_table)
R> out$assurance_plot

\end{CodeInput}
\begin{CodeOutput}
     n Assurance
1  100 0.5228078
2  110 0.5324414
3  120 0.5408288
4  130 0.5482139
5  140 0.5547789
6  150 0.5606632
\end{CodeOutput}
\end{CodeChunk}

\begin{figure}[t!]
\centering
\includegraphics[width = 8 cm]{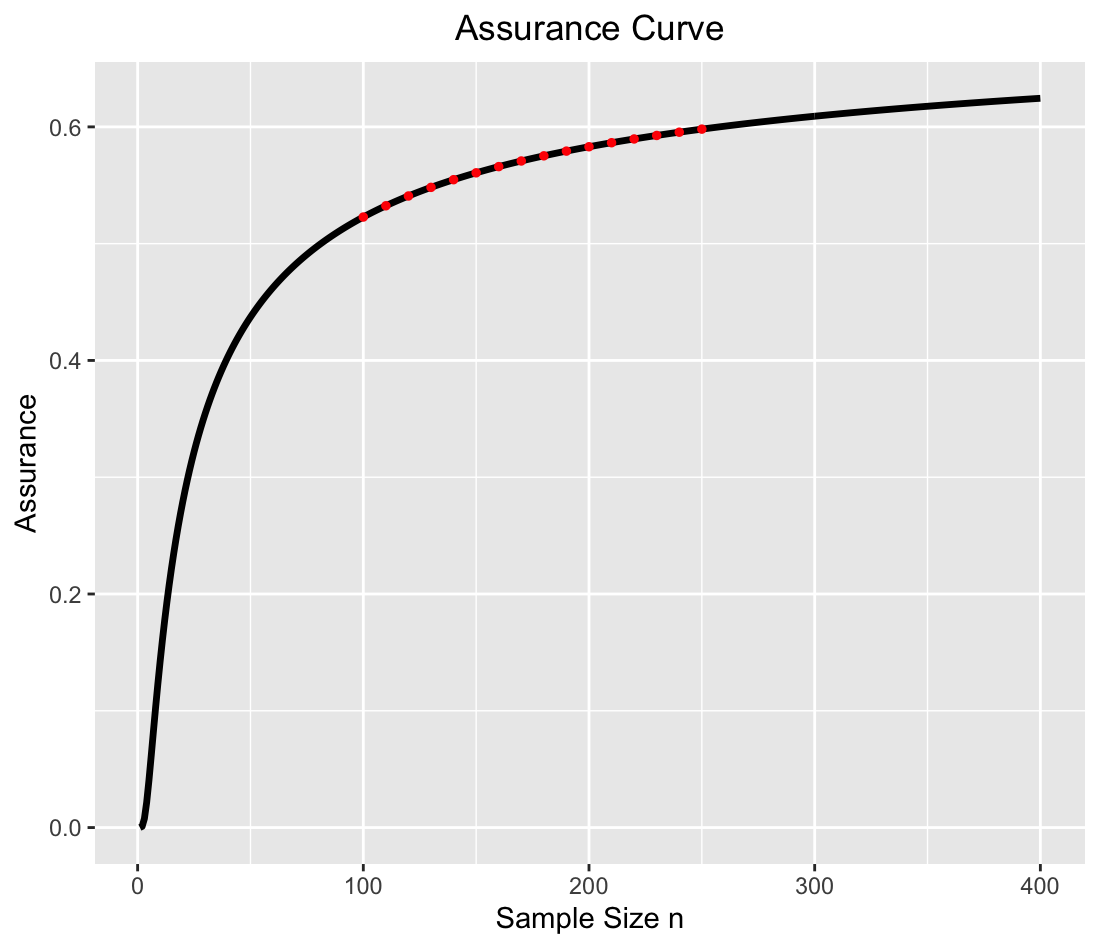}
\caption{\label{fig:assur_plot} Resulting assurance plot
with specific points passed in marked in red.}
\end{figure}

Running this block of code will return a table of assurance values 
and a graphical display of the assurance
curve, shown in Figure~\ref{fig:assur_plot}.  
The first six rows of the table are reported in the 
outputs. 

There are a few points worth noting from the
above code block as they are pertinent to the vast 
majority of functions contained within 
\pkg{bayesassurance}.  First, 
we are passing a vector of sample sizes
for \code{n} and saving the results as an 
arbitrary variable 
\code{out}.  The list of objects returned by the function 
is contingent on whether the user passes in a 
scalar or vector for \code{n}. 
If \code{n} is a scalar, this notifies the program that 
we only want to determine the assurance
for one particular sample size. When this is the case, 
\code{out} will only report a single assurance value with 
no plot. On the other hand, if a vector of sample sizes 
is passed in to \code{n}, as is the case for the code sample above,  
this suggests the user wants to determine the assurance for 
an array of sample sizes, and the function will produce both a table and 
an assurance curve showcasing the results. 
As long as \code{n} holds a length of at least two, 
\code{assurance\_nd\_na} will create a graphical display
of the assurance values for an array of sample sizes
surrounding those values of \code{n} that were
passed in, with specific points of interest 
labeled in red. Figure~\ref{fig:assur_plot} shows
the resulting assurance curve corresponding
to the code segment above. The graph is 
created using \pkg{ggplot2} which is 
imported into \pkg{bayesassurance}. 
Simply typing \code{out\$assurance\_table} and
\code{out\$assurance\_plot} will display the table
and plot respectively in this particular set
of examples.

\subsection{Special Case: Convergence with the Frequentist Setting}
Depending on how we define the parameters in 
\fct{assurance\_nd\_na}, we could demonstrate the 
direct relationship held between the Bayesian 
and classical settings, in which the frequentist power 
is no more than a special case of the generalized Bayesian solution. 
This can easily be seen when letting $n_d \rightarrow \infty$
and setting $n_a = 0$ in Equation~\eqref{eq:assurance}, i.e. 
defining a weak analysis stage prior and a strong design 
stage prior, resulting in
\begin{equation}
\label{eq:assurance_weakprior}
\Phi\Big(\sqrt{n}\frac{\Delta}{\sigma} + Z_\alpha\Big),
\end{equation}
which is equivalent to the frequentist power expression
that takes the form
\begin{equation}
\nonumber
1-\beta = P\Big(\bar{y} > \theta_0 +
 \frac{\sigma}{\sqrt{n}}Z_{1-\alpha}\Big) = 
\Phi\Big(\sqrt{n}\frac{\Delta}{\sigma} + Z_\alpha\Big),
\end{equation}
The following code chunk demonstrates this special case 
in \proglang{R} using the \fct{assurance\_nd\_na} function: 

\begin{CodeChunk}
\begin{CodeInput}

R> library(bayesassurance)

R> n <- seq(10, 250, 5)
R> n_a <- 1e-8
R> n_d <- 1e+8
R> theta_0 <- 0.15
R> theta_1 <- 0.25
R> sigsq <- 0.104

R> out <- assurance_nd_na(n = n, n_a = n_a,  n_d = n_d, 
theta_0 = theta_0, theta_1 = theta_1, sigsq = sigsq, 
alt = "greater", alpha = 0.05)

R> head(out$assurance_table)
R> out$assurance_plot

\end{CodeInput}
\begin{CodeOutput}
     n Assurance
1   10 0.2532578
2   15 0.3285602
3   20 0.3981637
4   25 0.4623880
5   30 0.5213579
6   35 0.5752063
\end{CodeOutput}
\end{CodeChunk}

The \pkg{bayesassurance} package includes 
a \fct{pwr\_freq} function that determines 
the statistical power of a study design 
given a set of fixed parameter values that 
adhere to the closed-form solution of power 
and sample size. Continuing with the one-sided
case, the solution is given by 
\begin{equation}
\label{eq:power_func}
1-\beta = P\Big(\bar{y} > \theta_0 +
 \frac{\sigma}{\sqrt{n}}Z_{1-\alpha}\Big) = 
\Phi\Big(\sqrt{n}\frac{\Delta}{\sigma} + Z_\alpha\Big),
\end{equation}
where $\Delta = \theta_1 - \theta_0$ 
denotes the critical difference and $\Phi$ denotes 
the cumulative distribution function of the standard
normal. Note this formula is equivalent to the 
special case of the assurance definition expressed 
in Equation~\eqref{eq:assurance_weakprior}.

To execute \fct{pwr\_freq}, the following set of parameters
need to be specified: 
\begin{enumerate}
 \item{\code{n}: sample size (either scalar or vector)}
 \item{\code{theta_0}: initial value specified in the null 
 hypothesis; typically provided by the client}
 \item{\code{theta_1}: alternative value to test against the 
 initial value; serves as a threshold in determining whether
 the null is to be rejected or not}
 \item{\code{alt}: specifies alternative test case, 
where \code{alt = "greater"} tests if 
$\theta_1 > \theta_0$,\code{alt = "less"} tests if 
$\theta_1 < \theta_0$, and \code{alt = "two.sided"}
performs a two-sided test for 
$\theta_1 \neq \theta_0$.  By default, 
\code{alt = "greater"}}.
 \item{\code{sigsq}: known variance}
 \item{\code{alpha}: significance level of test}
\end{enumerate}
As a simple example, consider the following code chunk
that directly runs \fct{pwr\_freq} through specifying
the above parameters and loading in \pkg{bayesassurance}: 
\begin{CodeChunk}
\begin{CodeInput}

R> library(bayesassurance)
R> pwr_freq(n = 20, theta_0 = 0.15, theta_1 = 0.35, sigsq = 0.30, 
		alt = "greater", alpha = 0.05)

\end{CodeInput}
\end{CodeChunk}
\begin{CodeOutput}
"Power: 0.495"
\end{CodeOutput}
Running this simply returns the assurance printed
as a statement since we are only evaluating one
sample size, \code{n = 20}. Now consider the next 
segment of code.
\begin{CodeChunk}
\begin{CodeInput}

R> library(bayesassurance)
R> n <- seq(10, 250, 5)
R> out <- pwr_freq(n = n, theta_0 = 0.15, theta_1 = 0.25, sigsq = 0.104, 
alt = "greater", alpha = 0.05)

R> head(out$pwr_table)
R> out$pwr_plot

\end{CodeInput}
\begin{CodeOutput}
     n     Power
1   10 0.2532578
2   15 0.3285602
3   20 0.3981637
4   25 0.4623880
5   30 0.5213579
6   35 0.5752063
\end{CodeOutput}
\end{CodeChunk}
This above code produces the exact same results as the
previous \fct{assurance\_nd\_na} example where we assigned
a weak analysis prior and a strong design prior to demonstrate
the overlapping behaviors that occur between
the two frameworks.


\begin{figure}%
    \centering
    \subfloat[\centering Power curve]{{\includegraphics[width=6cm]{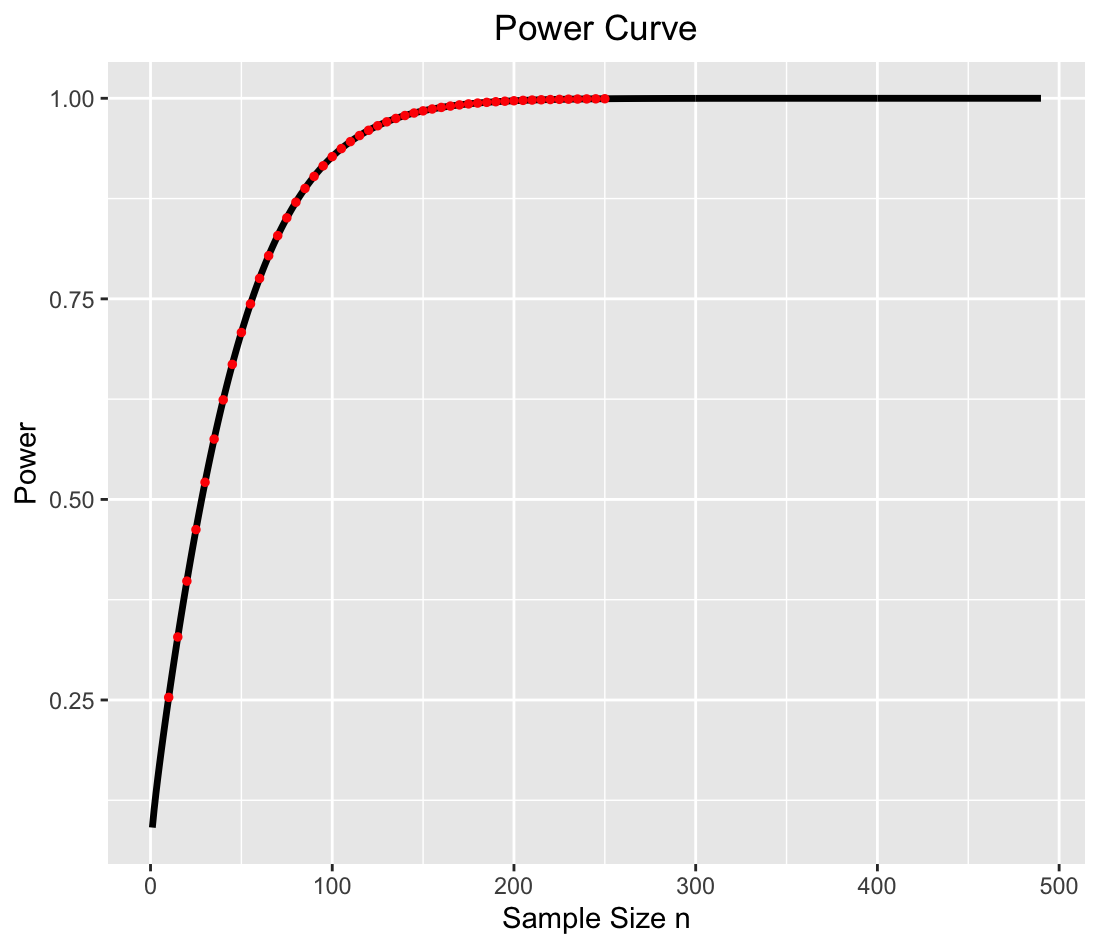} }}%
    \qquad
    \subfloat[\centering Assurance curve]{{\includegraphics[width=6cm]{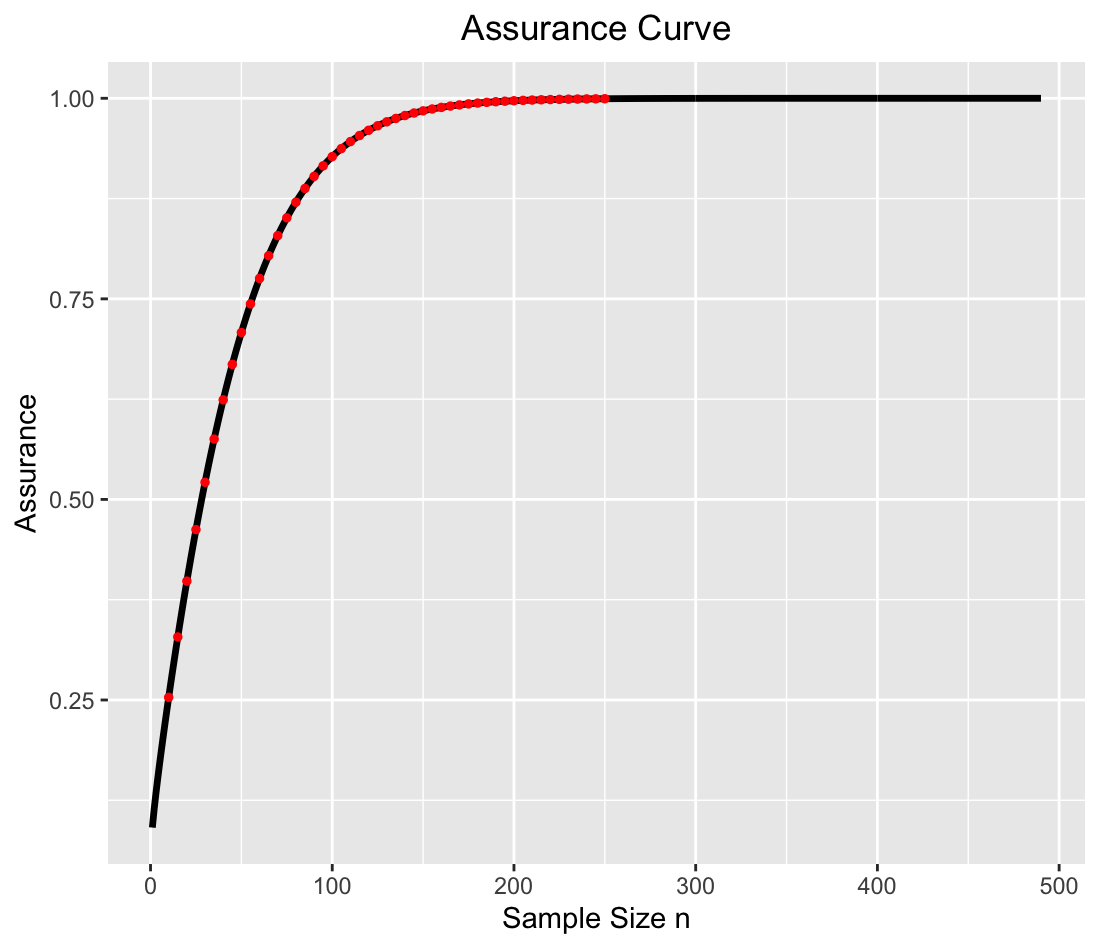} }}%
    \caption{Resulting power and assurance curve when weak analysis priors and strong design priors and enforced.}%
    \label{fig:pwr_assur_curve}%
\end{figure}

\section{Simulation-Based Functions Using Conjugate Linear Models}\label{sec:simform}

Hence forth, we focus on computing 
assurance through simulation-based means 
and highlight the commonplace scenario in sample 
size analysis where closed-form solutions 
are unavailable. In Sections~\ref{sec:simform} 
and~\ref{sec:othermethods}, we extend upon the 
two-stage design structure briefly mentioned in 
Section~\ref{sec:closedformassurance} and 
discuss how the design and analysis objectives are 
constructed based on the sample size criteria. 
Section~\ref{sec:simform}, in particular, draws attention to
simulations set within the conjugate Bayesian linear model
framework, while Section 4 addresses 
conditions delineated in other settings.

Each simulation-based function is characterized by a well-defined 
objective that we seek to evaluate the tenability of. This takes place 
in the analysis stage. The functions take an iterative approach that 
alternates between generating a dataset in the design stage and 
evaluating whether or not the dataset satisfies the analysis stage 
criteria. The assurance equates to the proportion of datasets 
that meet the objective. 

To elaborate on this idea in greater detail, we start in the analysis stage.
Consider a set of $n$ observations denoted by 
$y = (y_1, y_2, \cdots, y_n)^{\top}$ that are to be collected
in the presence of $p$ controlled explanatory variables,
$x_1, x_2, \cdots, x_p$. Specifically,
\[y = X_n\beta + \epsilon_n,\] 
where $X_n$ is an $n \times p$ design matrix with the $i$-th row 
corresponding to $x_i$, and $\epsilon_n \sim N(0, \sigma^2V_n)$, 
where $V_n$ is a known $n \times n$ correlation matrix. 
A conjugate Bayesian linear regression model specifies the
joint distribution of parameters $\{\beta, \sigma^2\}$
and $y$ as 
\[
IG(\sigma^2|a_{\sigma}, b_{\sigma}) \times N(\beta|\mu_{\beta}^{(a)},
\sigma^2V_{\beta}^{(a)}) \times N(y | X_n\beta, \sigma^2V_n),
\]
where superscripts $(a)$ indicate parameters in the analysis stage. 
The overarching objective is finding 
the assurance of the realized 
data favoring $H: u^{\top}\beta > C$, where $u$ is a 
$p \times 1$ vector of fixed contrasts and $C$ is a known constant. 
Inference proceeds from the posterior 
distribution given by 
\begin{equation}
\label{eq:post_dist}
p(\beta, \sigma^2|y) = IG(\sigma^2|a_{\sigma}^*, b_{\sigma}^*)
\times N(\beta|M_nm_n, \sigma^2M_n),
\end{equation}
where $a^*_{\sigma} = a_{\sigma}+\frac{n}{2}$, $b^*_{\sigma} = b_{\sigma} +
\frac{1}{2}\left\{\mu_{\beta}^{\top}V_{\beta}^{-1}\mu_{\beta} +
y_n^{\top}V_n^{-1}y - m_n^{\top}M_nm_n\right\}$, $M_n^{-1} =
V_{\beta}^{-1(a)} + X_n^{\top}V_n^{-1}X_n$ and $m_n =
V_{\beta}^{-1(a)}\mu_{\beta}^{(a)} + X_n^{\top}V_n^{-1}y$.
The posterior distribution helps shape our
analysis objective. 

If $\sigma^2$ is known and fixed, then the posterior distribution of 
$\beta$ is $\displaystyle p(\beta | \sigma^2, y) = N(\beta | 
M_nm_n, \sigma^2M_n)$ shown
in the Equation~\eqref{eq:post_dist}. 
We use the posterior components
of $\beta$ to evaluate $H: u^{\top}\beta > C$,
where standardization leads to
\begin{equation} \label{eq1}                                                                    
\left.\frac{u^{\top}\beta - u^{\top}M_nm_n}{\sigma \sqrt{u^{\top}M_nu}} \right| \sigma^2, y \sim N(0,1)\;.
\end{equation}
Hence, to assess the tenability of $H: u^{\top}\beta > C$, we decide 
in favor of $H$ if the observed data belongs in the set
\begin{align*}
    A_{\alpha}(u, \beta, C) &= \left\{y: P\left(u^{\top}\beta
    \leq C | y\right)  < \alpha\right\} = \left\{y: \Phi
    \left(\frac{C - u^{\top}M_n m_n}{\sigma \sqrt{u^{\top}M_nu}}\right)
    < \alpha \right\}.
\end{align*}
This defines our analysis objective, which we 
will monitor within each sample iteration. 
Sample generation is taken to account for in the design stage 
discussed in the next section. 

In the design stage, the goal is to seek a 
sample size $n$ such that the
analysis objective is met 100$\delta \%$ of the time,
where $\delta$ is the assurance.
This step requires determining the marginal 
distribution of $y$, which is assigned a separate set 
of priors to quantify our belief about the population 
from which the sample will be taken. Hence, the marginal 
of $y$ under the design priors will be derived from 
\begin{align}\label{eq: ohagan_stevens_linear_regression_design_priors}
  y &=  X_n\beta + e_n; \quad e_n \sim N(0, \sigma^2 V_n)\;;\quad \beta = \mu_{\beta}^{(d)} + \omega; \quad \omega \sim N(0, \sigma^2 V_{\beta}^{(d)})\;,
\end{align}
where $\beta\sim N(\mu_{\beta}^{(d)},\sigma^2 V_{\beta}^{(d)})$ is the design prior on $\beta$ and $(d)$ denotes parameters
in the design stage. Substituting the equation for $\beta$ into the equation for $y$ 
gives $y = X\mu_{\beta}^{(d)} + (X\omega + e_n)$ and, hence, $y \sim N\left(X\mu_{\beta}^{(d)}, \sigma^2V_n^{*}\right)$, where $V_n^{\ast} = \left(XV_{\beta}^{(d)}X^{\top} + V_n\right)$. 

To summarize our simulation strategy for estimating
the Bayesian assurance, we fix sample size $n$ and generate
a sequence of $J$ datasets $y^{(1)}, y^{(2)}, ..., y^{(J)}$.
A Monte Carlo estimate of the Bayesian assurance is
computationally determined as 
\[
\hat{\delta}(n) = \frac{1}{J}\sum_{j=1}^J \mathbb{I}\left(\left\{y^{(j)}:
    \Phi \left(\frac{C - u^{\top}M_n^{(j)}m_n^{(j)}}{\sigma
    \sqrt{u^{\top}M_n^{(j)}u}}\right) < \alpha \right\}\right)\;,
\]
where $\mathbb{I}(\cdot)$ is the indicator function of the
event in its argument, $M_n^{(j)}$ and $m_n^{(j)}$ are the
values of $M_n$ and $m_n$ computed from $y^{(j)}$.


\subsection{Assurance Computation with Known Variance}
\label{sec:assur_knownvar}
The simulation-based function, 
\fct{bayes\_sim}, determines the assurance
within the context of conjugate Bayesian linear regression models
assuming known variance, $\sigma^2$.
The execution of \fct{bayes\_sim} is straightforward. 
An important attribute is that users are not required to 
provide their own design matrix,
$X_n$, when executing \fct{bayes\_sim}.
The algorithm automatically accommodates 
for the case, \code{Xn = NULL}, using the 
built-in function, \fct{gen\_Xn}, which constructs appropriate 
design matrices based on entered sample size(s). 
Section~\ref{sec:tools} discusses design matrix
generators in greater detail. 
Setting \code{Xn = NULL} facilitates calculation
of assurances across
a vector of sample sizes, where the function sequentially 
updates the design matrix for each unique sample
size undergoing evaluation. 

Implementing \fct{bayes\_sim} requires defining the following 
set of parameters: 
\begin{enumerate}
\item{\code{n}: Sample size (either vector or scalar). 
If vector, each value corresponds to a separate study design. }
\item{\code{p}: Number of explanatory variables being considered.
Also denotes the column dimension of design matrix \code{Xn}. 
If \code{Xn = NULL}, \code{p} must be specified for the function to assign a default design matrix for \code{Xn}.}
\item{\code{u}: a scalar or vector included in the expression to be evaluated, e.g. $u^{\top}\beta > C$, where $\beta$ is an unknown parameter that is to be estimated.}
\item{\code{C}: constant to be compared to}
\item{\code{Xn}: design matrix characterizing the
observations given by the normal linear regression model 
$y_n = X_n\beta + \epsilon_n$, where $\epsilon_n \sim 
N(0, \sigma^2V_n)$. See above description for details. 
Default \code{Xn} is an $np \times p$ matrix comprised of $n \times 1$ 
ones vectors that run across the diagonal of the matrix.}
\item{\code{Vbeta_d}: correlation matrix that characterizes prior
information on $\beta$ in the design stage, i.e. $\beta \sim
N(\mu_{\beta}^{(d)}, \sigma^2 V_{\beta}^{(d)})$.}
\item{\code{Vbeta_a_inv}: inverse-correlation matrix that
characterizes prior information on $\beta$ in the analysis stage, i.e. 
$\beta \sim N(\mu_{\beta}^{(a)}, \sigma^2V_{\beta}^{(a)})$. 
The inverse is passed in for computation efficiency,
i.e. $V_{\beta}^{-1(a)}$.}
\item{\code{Vn}: an $n \times n$ correlation matrix for the marginal
distribution of the sample data $y_n$. Takes on an identity matrix when
set to \code{NULL}.}
\item{\code{sigsq}: a known and fixed constant preceding all
correlation matrices \code{Vn}, \code{Vbeta_d} and
\code{Vbeta_a_inv}.}
\item{\code{mu_beta_d}: design stage mean, $\mu_{\beta}^{(d)}$}
\item{\code{mu_beta_a}: analysis stage mean, $\mu_{\beta}^{(a)}$}
\item{\code{alpha}: specifies alternative test case, where 
\code{alt = "greater"} tests if $u^{\top}\beta > C$,
\code{alt = "less"} tests if $u^{\top}\beta < C$, and 
\code{alt = "two.sided"} performs a two-sided test for
$u^{\top}\beta \neq C$. 
By default, \code{alt = "greater"}.}
\item{\code{alpha}: significance level}
\item{\code{mc_iter}: number of MC samples evaluated under the analysis objective}
\end{enumerate}

\subsubsection{Example 1: Scalar Parameter}

The first example evaluates the 
tenability of $H: u^{\top}\beta > C$ in the case when
$\beta$ is a scalar. The following code segment
assigns a set of arbitrary values for the parameters of 
\fct{bayes\_sim} and saves the outputs as \code{assur\_vals}.
The first ten rows of the table is shown. 
\begin{CodeChunk}
\begin{CodeInput}
   
R> n <- seq(100, 300, 10)

R> assur_vals <- bayesassurance::bayes_sim(n, p = 1, u = 1,
      C = 0.15, Xn = NULL, Vbeta_d = 0, Vbeta_a_inv = 0,
      Vn = NULL, sigsq = 0.265, mu_beta_d = 0.25, mu_beta_a = 0,
      alt = "greater", alpha = 0.05, mc_iter = 5000)

R> head(assur_vals$assurance_table)
R> assur_vals$assurance_plot

\end{CodeInput}
\begin{CodeOutput}
   Observations per Group (n) Assurance
1                         100    0.6162
2                         110    0.6612
3                         120    0.6886
4                         130    0.7148
5                         140    0.7390
6                         150    0.7746
\end{CodeOutput}
\end{CodeChunk}

We emphasize a few important points in this 
rudimentary example. Assigning a vector of 
values for \code{n} indicates we are interested in assessing
the assurance for multiple study designs. Each unique value
passed into \code{n} corresponds to a separate balanced study design
containing that particular sample size for each of the $p$
groups undergoing assessment. In this example, 
setting \code{p = 1}, \code{u = 1} and \code{C = 0.15} 
implies we are evaluating the tenability of $H: \beta > 0.15$, 
where $\beta$ is a scalar.Furthermore, \code{Vbeta_d} and
\code{Vbeta_a_inv} are scalars to align 
with the dimension of $\beta$. A weak analysis prior
(\code{Vbeta_a_inv = 0}) and a strong design prior
(\code{Vbeta_d = 0}) are assigned to 
demonstrate the overlapping scenario taking place 
between the Bayesian and frequentist settings.
Section~\ref{sec:tools} revisits this example when
reviewing features that allow users to simultaneously
visualize the Bayesian and frequentist settings in a 
single window. 
Finally, \code{Xn} and \code{Vn} are set to \code{NULL} 
which means they will take on the default settings 
specified in the parameter descriptions above. 

\subsubsection{Example 2: Linear Contrasts}
\label{subsec:ex2}

In this example, we assume $\beta$ is a vector of
unknown components rather than a scalar. We refer to
the cost-effectiveness application discussed in~\cite{ohagan}
to demonstrate a real-world setting. The application 
considers a randomized clinical trial 
that compares the cost-effectiveness of two treatments.
The cost-effectiveness is evaluated using
a net monetary benefit measure expressed as
\[
\xi = K(\mu_2 - \mu_1) - (\gamma_2 - \gamma_1),
\]
where $\mu_1$ and $\mu_2$ respectively denote the efficacy
of treatments 1 and 2, and $\gamma_1$ and
$\gamma_2$ denote the costs. 
Hence, $\mu_2 - \mu_1$ and $\gamma_2 - \gamma_1$
correspond to the true differences in 
treatment efficacy and costs, respectively,
between Treatments 1 and 2. The threshold unit cost, 
$K$, represents the maximum price that a 
health care provider is willing to pay for a 
unit increase in efficacy. 

In this setting,we seek the tenability of $H: \xi > 0$, which 
if true, indicates that Treatment 2 is more
cost-effective than Treatment 1.To comply with
the conjugate linear model framework outlined
in Equation~\eqref{eq:post_dist}, we set 
$u = (-K, 1, K, -1)^{\top}$, $\beta = (\mu_1, \gamma_1, \mu_2, \gamma_2)^{\top}$, and $C = 0$, giving us 
an equivalent form of $\xi > 0$ expressed as 
$u^{\top}\beta > 0$. All other inputs of this application
were directly pulled from the paper.
The following code sets up the inputs
to be passed into \fct{bayes\_sim}. 

\begin{CodeChunk}
\begin{CodeInput}

R> n <- 285
R> p <- 4
R> K <- 20000 # threshold unit cost
R> C <- 0
R> u <- as.matrix(c(-K, 1, K, -1))
R> sigsq <- 4.04^2

## Assign mean parameters to analysis and design stage priors
R> mu_beta_d <- as.matrix(c(5, 6000, 6.5, 7200))
R> mu_beta_a <- as.matrix(rep(0, p))

## Assign correlation matrices (specified in paper) 
## to analysis and design stage priors
R> Vbeta_a_inv <- matrix(rep(0, p^2), nrow = p, ncol = p)
R> Vbeta_d <- (1 / sigsq) * matrix(c(4, 0, 3, 0, 0, 10^7, 0, 
    0, 3, 0, 4, 0, 0, 0, 0, 10^7), nrow = 4, ncol = 4)

R> tau1 <- tau2 <- 8700
R> sig <- sqrt(sigsq)
R> Vn <- matrix(0, nrow = n*p, ncol = n*p)
R> Vn[1:n, 1:n] <- diag(n)
R> Vn[(2*n - (n-1)):(2*n), (2*n - (n-1)):(2*n)] <- (tau1 / sig)^2 * diag(n)
R> Vn[(3*n - (n-1)):(3*n), (3*n - (n-1)):(3*n)] <- diag(n)
R> Vn[(4*n - (n-1)):(4*n), (4*n - (n-1)):(4*n)] <- (tau2 / sig)^2 * diag(n)

\end{CodeInput}
\end{CodeChunk}

The inputs specified above should result in 
an assurance of approximately 0.70 according to
~\cite{ohagan}. The \fct{bayes\_sim}
returns a similar value, demonstrating
that sampling from the posterior yields results similar
to those reported in the paper. 

\begin{CodeChunk}
\begin{CodeInput}

R> library(bayesassurance)

R> assur_vals <- bayes_sim(n = 285, p = 4,  u = as.matrix(c(-K, 1, K, -1)), 
   C = 0, Xn = NULL, Vbeta_d = Vbeta_d,Vbeta_a_inv = Vbeta_a_inv, 
   Vn = Vn, sigsq = 4.04^2, mu_beta_d = as.matrix(c(5, 6000, 6.5, 7200)), 
   mu_beta_a = as.matrix(rep(0, p)),  alt = "greater", alpha = 0.05, mc_iter = 10000)
   
R> assur_vals

\end{CodeInput}
\begin{CodeOutput}
## [1] "Assurance: 0.722"
\end{CodeOutput}
\end{CodeChunk}

\subsection{Assurance Computation in the Longitudinal Setting}
\label{sec:assur_longitudinal}
We demonstrate an additional feature
embedded in the function tailored to longitudinal data. 
In this setting, $n$ no longer refers to the number of subjects
but rather the number of repeated measures reported
for each subject assuming a balanced study design. 
Referring back to the linear regression model discussed
in the general framework, 
we can construct a longitudinal model that utilizes
this same linear regression form, where
$y_n = X_n\beta + \epsilon_n$.

Consider a group of subjects in a balanced longitudinal
study with the same number of repeated measures at 
equally-spaced time points. 
In the base case, in which time is treated as a linear term, 
subjects can be characterized by 
\[
y_{ij} = \alpha_{i} + \beta_{i}t_{ij} + \epsilon_{i}, 
\]
where $y_{ij}$ denotes the $j^{th}$ observation of subject
$i$ at time $t_{ij}$, $\alpha_{i}$ and $\beta_{i}$ respectively
denote the intercept and slope terms for subject $i$, 
and $\epsilon_{i}$ is an error term characterized by 
$\epsilon_i \sim N(0, \sigma^2_i)$. 

In a simple case with two subjects, we can individually 
express the observations as
\begin{align*}
y_{11} &= \alpha_{1} + \beta_{1}t_{11} + \epsilon_{1}\\
&\vdots\\
y_{1n} &= \alpha_{1} + \beta_{1}t_{1n} + \epsilon_{1}\\
y_{21} &= \alpha_{2} + \beta_{2}t_{21} + \epsilon_{2}\\
&\vdots\\
y_{2n} &= \alpha_{2} + \beta_{2}t_{2n} + \epsilon_{2},
\end{align*}
assuming that each subject contains $n$ observations. 
The model can also be expressed cohesively using matrices,
\begin{gather}
\label{eq:long_model}
 \underbrace{\begin{pmatrix} y_{11} \\ \vdots \\ y_{1n} \\ y_{21} \\ \vdots\\ y_{2n} \end{pmatrix}}_{y_n} =
   \underbrace{
   \begin{pmatrix}
   1 & 0 & t_{11} & 0\\
   \vdots & \vdots & \vdots & \vdots \\
   1 & 0 & t_{1n} & 0\\
   0 & 1 & 0 & t_{21}\\
   \vdots & \vdots & \vdots & \vdots \\
   0 & 1 & 0 & t_{2n}
   \end{pmatrix}
   }_{X_n}
   \underbrace{
      \begin{pmatrix}
         \alpha_{1}\\
         \alpha_{2}\\
         \beta_{1}\\
         \beta_{2}
      \end{pmatrix}
   }_{\beta}
   + 
   \underbrace{
   \begin{pmatrix}
   \epsilon_{1}\\
   \vdots\\
   \epsilon_{1}\\
   \epsilon_{2}\\
   \vdots\\
   \epsilon_{2}
   \end{pmatrix}
   }_{\epsilon_n}
\end{gather}
bringing us back to the linear model structure. If higher degrees
are to be considered for the time variable, such as the inclusion
of a quadratic term, the model would be altered to 
include additional covariate terms that can accommodate for 
these changes. In the two-subject case, 
incorporating a quadratic term for the time variable
in Equation~\eqref{eq:long_model} will result
in the model being modified as follows: 
\begin{gather}
\nonumber
 \underbrace{\begin{pmatrix} y_{11} \\ \vdots \\ y_{1n} \\ y_{21} \\ \vdots\\ y_{2n} \end{pmatrix}}_{y_n} =
   \underbrace{
   \begin{pmatrix}
   1 & 0 & t_{11} & 0 & t_{11}^2 & 0\\
   \vdots & \vdots & \vdots & \vdots & \vdots & \vdots\\
   1 & 0 & t_{1n} & 0 & t_{1n}^2  & 0\\
   0 & 1 & 0 & t_{21} & 0 & t_{21}^2 \\
   \vdots & \vdots & \vdots & \vdots & \vdots & \vdots \\
   0 & 1 & 0 & t_{2n} & 0 & t_{2n}^2
   \end{pmatrix}
   }_{X_n}
   \underbrace{
      \begin{pmatrix}
         \alpha_{1}\\
         \alpha_{2}\\
         \beta_{1}\\
         \beta_{2}\\
         \phi_{1}\\
         \phi_{2}
      \end{pmatrix}
   }_{\beta}
   + 
   \underbrace{
   \begin{pmatrix}
   \epsilon_{1}\\
   \vdots\\
   \epsilon_{1}\\
   \epsilon_{2}\\
   \vdots\\
   \epsilon_{2}
   \end{pmatrix}
   }_{\epsilon_n}
\end{gather}
In general, for $m$ subjects who each have
$n$ repeated measures, a one-unit increase in the degree of the
time-based covariate will
result in $m$ additional columns being added to the 
design matrix $X_n$ and $m$ additional rows added to the
$\beta$ vector. 

When working in the longitudinal setting,
additional parameters need to be specified
in the \fct{bayes\_sim} function.
These include
\begin{enumerate}
\item{\code{longitudinal}: logical that 
indicates the simulation will be based in a longitudinal setting. 
If \code{Xn = NULL}, the function will construct a design 
matrix using inputs that correspond to a balanced
longitudinal study design.}
\item{\code{ids}: vector of unique subject ids}
\item{\code{from}: start time of repeated measures for each subject}
\item{\code{to}: end time of repeated measures for each subject}
\item{\code{num_repeated_measures}: desired length of 
the repeated measures sequence.This should be a non-negative
number, will be rounded up otherwise if fractional.}
\item{\code{poly_degree}: degree of polynomial in longitudinal model, set to 1 by default.}
\end{enumerate}
By default, \code{longitudinal = FALSE} and \code{ids,
from}, and \code{to} are set to \code{NULL} when
working within the standard conjugate linear model. 
When \code{longitudinal = TRUE}, \code{n} takes on a 
different meaning as its value(s) correspond to the number 
of repeated measures for each subject rather than the 
total number of subjects in each group. 
When \code{longitudinal = TRUE} and \code{Xn = NULL},
\fct{bayes\_sim} implicitly relies on a design matrix generator,
\fct{gen\_Xn\_longitudinal}, that is specific to the longitudinal setting
to construct appropriate design matrices. 
Section~\ref{sec:tools} discusses
this in greater detail. 
%
%

\subsubsection{Example 3: Longitudinal Example}
\begin{figure}[t!]
\centering
\includegraphics[width = 8 cm]{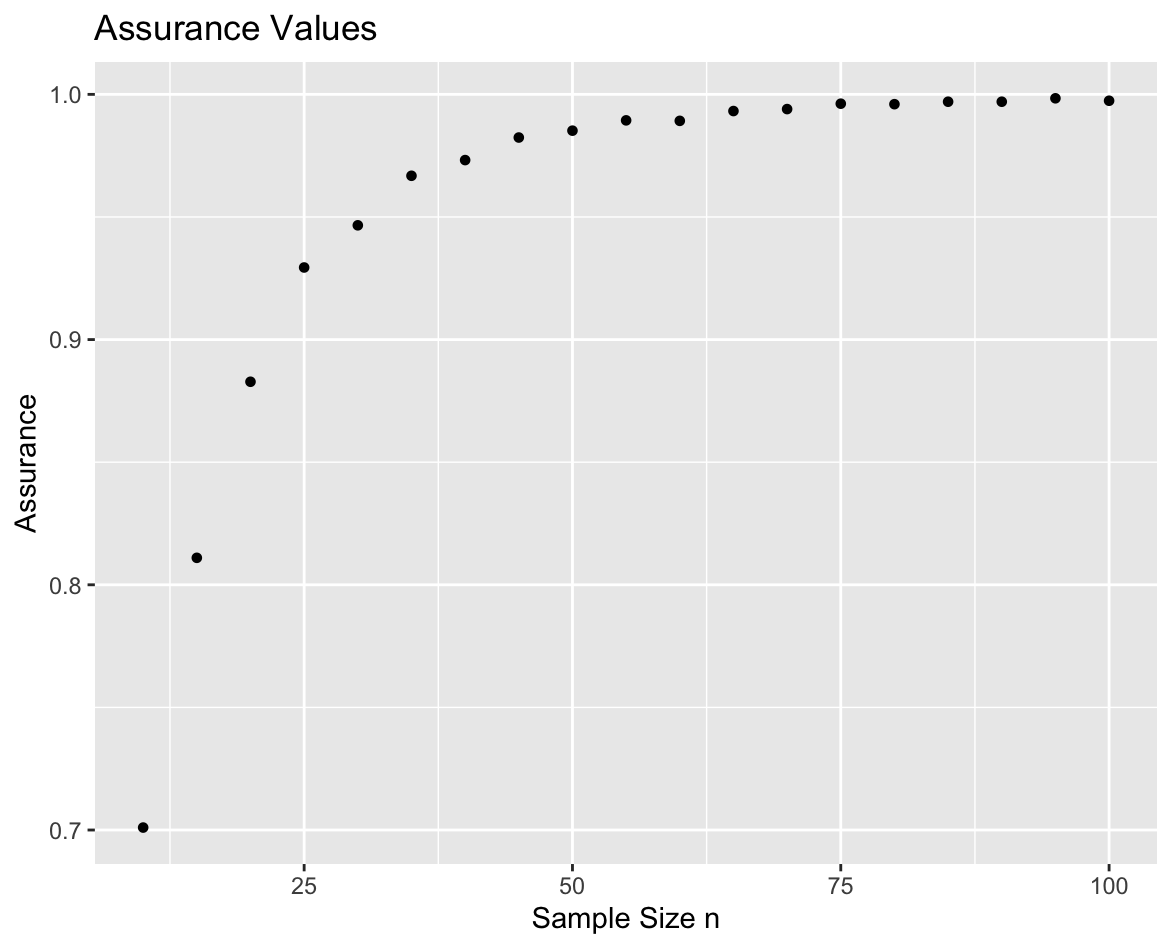}
\caption{\label{fig:ex3} Estimated assurance points for longitudinal example.}
\end{figure}

The following example uses similar parameter settings as 
the cost-effectiveness example we had previously discussed in
Section~\ref{subsec:ex2}, now with 
longitudinal specifications. We assume two subjects and
want to test whether the growth rate of subject 1
is different in comparison to subject 2. This could have either positive
or negative implications depending on the measurement scale. 
Figure~\ref{fig:ex3} displays the estimated assurance points
given the specifications. 


Assigning an appropriate linear contrast
lets us evaluate the tenability of an outcome. 
Let us consider the tenability of 
$u^{\top}\beta \neq C$ in this next example, where
$u = (1, -1, 1, -1)^{\top}$ and $C = 0$. The timepoints
are arbitrarily chosen to be 0 through 120- this
could be days, months, or years depending on the context
of the problem. The number of repeated measurements per subject
to be tested includes values 10 through 100 in increments of 5. 
This indicates that we are evaluating
the assurance for 19 study designs in total. $n = 10$ 
divides the specified time interval 
into 10 evenly-spaced timepoints between 0 and 120.

For a more complicated study design comprised of more
than two subjects that are divided into two treatment groups, 
consider testing if the mean growth rate is higher in 
the first treatment group than that of the second, 
e.g. if we have three subjects per treatment group, 
the linear contrast would be set as 
$u = (0, 0, 0, 0, 0, 0, 1/3, 1/3, 1/3, -1/3, -1/3, -1/3)^{\top}$.

\begin{CodeChunk}
\begin{CodeInput}
R> n <- seq(10, 100, 5)
R> ids <- c(1,2)
R> Vbeta_a_inv <- matrix(rep(0, 16), nrow = 4, ncol = 4)
R> sigsq = 100
R> Vbeta_d <- (1 / sigsq) * matrix(c(4, 0, 3, 0, 0, 6, 0, 0, 3, 0, 4, 0, 0, 0, 0, 6), 
                                nrow = 4, ncol = 4)

R> assur_out <- bayes_sim(n = n, p = NULL, u = c(1, -1, 1, -1), C = 0, Xn = NULL,
                       Vbeta_d = Vbeta_d, Vbeta_a_inv = Vbeta_a_inv,
                       Vn = NULL, sigsq = 100,
                       mu_beta_d = as.matrix(c(5, 6.5, 62, 84)),
                       mu_beta_a = as.matrix(rep(0, 4)), mc_iter = 5000,
                       alt = "two.sided", alpha = 0.05, longitudinal = TRUE, ids = ids,
                       from = 10, to = 120)
                       
R> head(assur_out$assurance_table)
R> assur_out$assurance_plot                 
\end{CodeInput}
\begin{CodeOutput}
  Observations per Group (n) Assurance
1                         10    0.6922
2                         15    0.8056
3                         20    0.8810
4                         25    0.9244
5                         30    0.9478
6                         35    0.9626
\end{CodeOutput}
\end{CodeChunk}

\subsection{Assurance Computation with Unknown Variance}
\label{sec:assur_unknownvar}
The \fct{bayes\_sim\_unknownvar} function operates similarly to 
\fct{bayes\_sim} but is used when the variance, 
$\sigma^2$, is unknown. 
In the unknown variance setting, priors are 
assigned to both $\beta$ and 
$\sigma^2$ in the analysis stage such that 
$\beta | \sigma^2 \sim N(\mu_{\beta}^{(a)},
\sigma^2V_{\beta}^{(a)})$ and $\sigma^{2} \sim IG(a^{(a)}, b^{(a)})$, 
where superscripts $(a)$ indicate analysis priors.
Determining the posterior distribution of $\sigma^{2}$ requires integrating out $\beta$ from the joint posterior distribution of $\{\beta, \sigma^2\}$, yielding 
\begin{equation}\label{eq: postbeta}
   \begin{split}
   p(\sigma^2| y_n) &\propto IG(\sigma^2| a^{(a)},b^{(a)}) \times \int
   N(\beta | \mu_{\beta}, \sigma^2 V_{\beta}) \times N(y_n | X\beta,
   \sigma^2 V_{n}) d\beta\\ &\propto \left(\frac{1}{\sigma^2}\right)^{a^{(a)}+
   \frac{n}{2} + 1}\exp\left\{-\frac{1}{\sigma^2}\left(b^{(a)} +
   \frac{c^{\ast}}{2}\right)\right\}\;.
   \end{split}
\end{equation}
Hence, $p(\sigma^2| y_n) =
IG\left(\sigma^2| a^{\ast}, b^{\ast}\right)$, where $a^{\ast} = a^{(a)} + \frac{n}{2}$ and $b^{\ast} = b^{(a)} + \frac{c^{\ast}}{2}$, where 
$c^{\ast} = b^{(a)} + \frac{1}{2}\Big(\mu_{\beta}^{\top (a)} V_{\beta}^{-1(a)}\mu_{\beta}^{(a)} +y_n^{\top}V_n^{-1}y_n -  m_n^{\top}M_nm_n\Big)$.
%

Recall the design stage aims to identify a minimum 
sample size that is needed to attain the assurance level specified by the investigator. 
We will need the marginal distribution of $y_n$ with priors placed on both $\beta$ and $\sigma^2$. We denote these design priors as $\beta^{(d)}$ and $\sigma^{2(d)}$. 
With $\sigma^{2(d)}$ now treated as an unknown parameter, the marginal distribution of $y_n$, given $\sigma^{2(d)}$, under the design prior is derived from $y_n =  X_n\beta^{(d)} + e_n$, $ e_n \sim N(0, \sigma^{2(d)} V_n)$, $\beta^{(d)} = \mu_{\beta}^{(d)} + \omega; \quad \omega \sim  N(0, \sigma^{2(d)} V_{\beta}^{(d)})$, where $\beta^{(d)}\sim N(\mu_{\beta}^{(d)},\sigma^{2(d)} V_{\beta}^{(d)})$
and $\sigma^{2(d)} \sim IG(a^{(d)}, b^{(d)})$. Substituting $\beta^{(d)}$ into
$y_n$ gives us $y_n = X_n\mu_{\beta}^{(d)} + (X_n\omega + e_n)$ such that $X_n\omega + e_n \sim N(0, \sigma^{2(d)}(V_n + X_n V_{\beta}^{(d)} X_n^{\top}))$.
The marginal distribution of $p(y_n| \sigma^{2(d)})$ is
\begin{equation}\label{eq:marg_yn}
   y_n | \sigma^{2(d)} \sim N(X_n \mu_{\beta}^{(d)}, \sigma^{2(d)} V_n^{*}); \quad
   V_n^{*} = X_n V_{\beta}^{(d)} X_n^{\top} + V_n.
\end{equation}
Equation~\eqref{eq:marg_yn} specifies our data generation model for ascertaining sample size.

To summarize, in the design stage, we draw $\sigma^{2(d)}$ from
$IG(a_\sigma^{(d)}, b_\sigma^{(d)})$ and generate the data from our sampling distribution from Equation~\eqref{eq:marg_yn}. For each data set,we perform Bayesian inference for $\beta$ and $\sigma^2$. Here, we draw $J$ samples of $\beta$ and $\sigma^2$ from their respective posterior distributions and compute the proportion of  these $J$ samples that satisfy $u^{\top} \beta_j > C$. If the proportion exceeds a certain threshold $1-\alpha$, then the analysis objective is met for that dataset. The above steps for the design and analysis stage are repeated for $R$ datasets and the proportion of the $R$ datasets that meet the analysis objective corresponds to the Bayesian assurance.

The \fct{bayes\_sim\_unknownvar} function requires the same
set of parameters as \fct{bayes\_sim} in addition to the following: 
\begin{enumerate}
\item{\code{R}: number of datasets generated used to check if the analysis objective is met. The proportion of those \code{R} datasets meeting the analysis objective corresponds to the estimated assurance.}
\item{\code{a\_sig\_d,b\_sig\_d}: shape and scale parameters of the inverse gamma distribution where variance $\sigma^2$
is sampled from in the design stage.}
\item{\code{a\_sig\_a,b\_sig\_a}: shape and scale parameters of the inverse gamma distribution where variance $\sigma^2$
is sampled from in the analysis stage.}
\end{enumerate}

\subsubsection{Example 4: Unknown Variance}
Referring back to the cost-effectiveness application presented in 
\cite{inoue}, we modify the inputs from \fct{bayes\_sim} 
and perform the simulation assuming unknown variance $\sigma^2$
using \fct{bayes\_sim\_unknownvar}. We implement the following code
for a single sample size $n = 285$ with assigned correlation
matrices mentioned in the paper. 
\begin{CodeChunk}
\begin{CodeInput}
R> library(bayesassurance)
                                
R> n <- 285
R> p <- 4 
R> K <- 20000
R> C <- 0
R> u <- as.matrix(c(-K, 1, K, -1))
R> Vbeta_a_inv <- matrix(rep(0, p^2), nrow = p, ncol = p)
R> sigsq <- 4.04^2
R> tau1 <- tau2 <- 8700

## manually-assigned correlation matrices
R> Vn <- matrix(0, nrow = n*p, ncol = n*p)
R> Vn[1:n, 1:n] <- diag(n)
R> Vn[(2*n - (n-1)):(2*n), (2*n - (n-1)):(2*n)] <- (tau1 / sig)^2 * diag(n)
R> Vn[(3*n - (n-1)):(3*n), (3*n - (n-1)):(3*n)] <- diag(n)
R> Vn[(4*n - (n-1)):(4*n), (4*n - (n-1)):(4*n)] <- (tau2 / sig)^2 * diag(n)

R> Vbeta_d <- (1 / sigsq) * matrix(c(4, 0, 3, 0, 0, 10^7, 0, 0, 3, 0, 4, 0, 
0, 0, 0, 10^7),
nrow = 4, ncol = 4)

R> mu_beta_d <- as.matrix(c(5, 6000, 6.5, 7200))
R> mu_beta_a <- as.matrix(rep(0, p))
R> alpha <- 0.05
R> epsilon <- 10e-7
R> a_sig_d <- (sigsq / epsilon) + 2
R> b_sig_d <- sigsq * (a_sig_d - 1)
R> a_sig_a <- -p / 2
R> b_sig_a <- 0
R> R <- 150

R> assur_out <- bayes_sim_unknownvar(n = n, p = 4, u = as.matrix(c(-K, 1, K, -1)), 
C = 0, R = 150,Xn = NULL, Vn = Vn, Vbeta_d = Vbeta_d, 
Vbeta_a_inv = Vbeta_a_inv, mu_beta_d = mu_beta_d,
mu_beta_a = mu_beta_a, a_sig_a = a_sig_a,
b_sig_a = b_sig_a, a_sig_d = a_sig_d,b_sig_d = b_sig_d, 
		alt = "two.sided", alpha = 0.05, mc_iter = 5000)
R> assur_out$assur_val
\end{CodeInput}
\begin{CodeOutput}
[1] "Assurance: 0.607"
\end{CodeOutput}
\end{CodeChunk}

\subsection{Assurance Computation for Unbalanced Study Designs}
\label{sec:assur_unbalanced}

The \fct{bayes\_sim\_unbalanced} function operates similarly to
\fct{bayes\_sim} in Section~\ref{sec:assur_knownvar}
but estimates the assurance of attaining $u^{\top}\beta > C$ 
specifically in an unbalanced design setting.
Users provide two sets of sample sizes of equal length,
whose corresponding pairs are considered for  
each study design case. The sample sizes need not
be equal to one another, allowing for unbalanced designs.
This is unlike \fct{bayes\_sim} that strictly 
determines assurance for balanced cases, where
users specify a single set of sample sizes
whose individual entries correspond 
to a distinct trial comprised of an equal number of observations
across all explanatory variables.  
The \fct{bayes\_sim\_unbalanced} function provides 
a higher degree of flexibility for designing unbalanced studies
and offers a more advanced visualization feature. Users
have the option of viewing assurance as a 3-D contour 
plot and assess how the assurance behaves
across varying combinations of the two 
sets of sample sizes that run along the 
$x$ and $y$ axes.

The \fct{bayes\_sim\_unbalanced} function is similar to
\fct{bayes\_sim} in terms of parameter specifications with a few
exceptions. Parameters unique to \fct{bayes\_sim\_unbalanced} 
are summarized below:
\begin{enumerate}
\item{\code{n1}: first sample size (either vector or scalar). }
\item{\code{n2}: second sample size (either vector or scalar). }
\item{\code{repeats}: an integer value denoting the number of 
times \code{c(n1, n2)} is accounted for; applicable for study designs 
that consider an even number of explanatory variables greater
than two and whose sample sizes correspond to those specified in
\code{n1} and \code{n2}.  By default, \code{repeats = 1}. See
Example 6 below. }
\item{\code{surface_plot}: logical parameter that 
indicates whether a contour plot is to be constructed. When set to \code{TRUE}, and \code{n1} and \code{n2} are vectors,
a contour plot (i.e. heat map) showcasing assurances 
obtained for unique combinations of \code{n1} and \code{n2} is produced. }
\end{enumerate}
As in \code{bayes\_sim}, it is recommended that users 
set \code{Xn = NULL} to facilitate the automatic 
construction of appropriate design matrices that best aligns with
the conjugate linear model described in beginning of 
Section~\ref{sec:simform}. Recall that every unique sample
size (or sample size pair) passed in corresponds to 
a separate study that requires a separate design matrix. 
Should users choose to provide their own design matrix, 
it is advised that they evaluate the assurance for
one study design at a time,
in which a single design matrix is passed
into \code{Xn} along with scalar values assigned for the
sample size parameter(s). 

Saved outputs from executing the function include
\begin{enumerate}
\item{\code{assurance_table}: table of sample size 
and corresponding assurance values}
\item{\code{contourplot}: contour map of assurance values 
if \code{surface.plot = TRUE}}
\item{\code{mc_samples}: number of Monte Carlo 
samples that were generated for evaluation}
\end{enumerate}

\subsubsection{Example 5: Unbalanced Assurance Computation with Surface Plot}

The following code provides a basic example of how
\fct{bayes\_sim\_unbalanced} is executed. It is important to check 
that the parameters passed in are appropriate in dimensions, e.g. 
\code{mu_beta_a} and \code{mu_beta_d} should each 
contain the same length as that of \code{u}, and the length of 
\code{u} should be equal to the row and column
dimensions of \code{Vbeta\_d} and \code{Vbeta\_a\_inv}.  

A table of assurance values is printed simply
by calling \code{assur_out$assurance_table}, which contains
the exact assurance values corresponding to each
sample size pair. The contour 
plot, shown in Figure~\ref{fig:ex5}, is displayed
using \code{assur_out$contourplot}, and offers a visual
depiction of how the assurance varies across unique
combinations of \code{n1} and \code{n2}.  Areas 
with lighter shades denote higher assurance levels.
No discernible patterns or trends are observed based on  
the random behavior of the plot and the proximity of values
reported in the table as the inputs were arbitrarily chosen
with no context. The next example implements the
function in a real-world setting that offers more sensible
results.
\begin{CodeChunk}
\begin{CodeInput}
R> library(bayesassurance)

R> n1 <- seq(20, 75, 5)
R> n2 <- seq(50, 160, 10)


R> assur_out <- bayes_sim_unbalanced(n1 = n1, n2 = n2, repeats = 1, u = c(1, -1),
	  C = 0, Xn = NULL, Vbeta_d = matrix(c(50, 0, 0, 10),nrow = 2, ncol = 2),
	  Vbeta_a_inv = matrix(rep(0, 4), nrow = 2, ncol = 2),
	  Vn = NULL, sigsq = 100,  mu_beta_d = c(1.17, 1.25),
	  mu_beta_a = c(0, 0), alt = "two.sided", alpha = 0.05, mc_iter = 5000,
	  surface_plot = TRUE)
				
R> head(assur_out$assurance_table)
R> assur_out$contourplot 

\end{CodeInput}
\begin{CodeOutput}
  n1  n2 Assurance
1 20  50    0.9504
2 25  60    0.9584
3 30  70    0.9508
4 35  80    0.9616
5 40  90    0.9624
6 45 100    0.9634
\end{CodeOutput}
\end{CodeChunk}

\begin{figure}[t!]
\centering
\includegraphics[width = 10 cm]{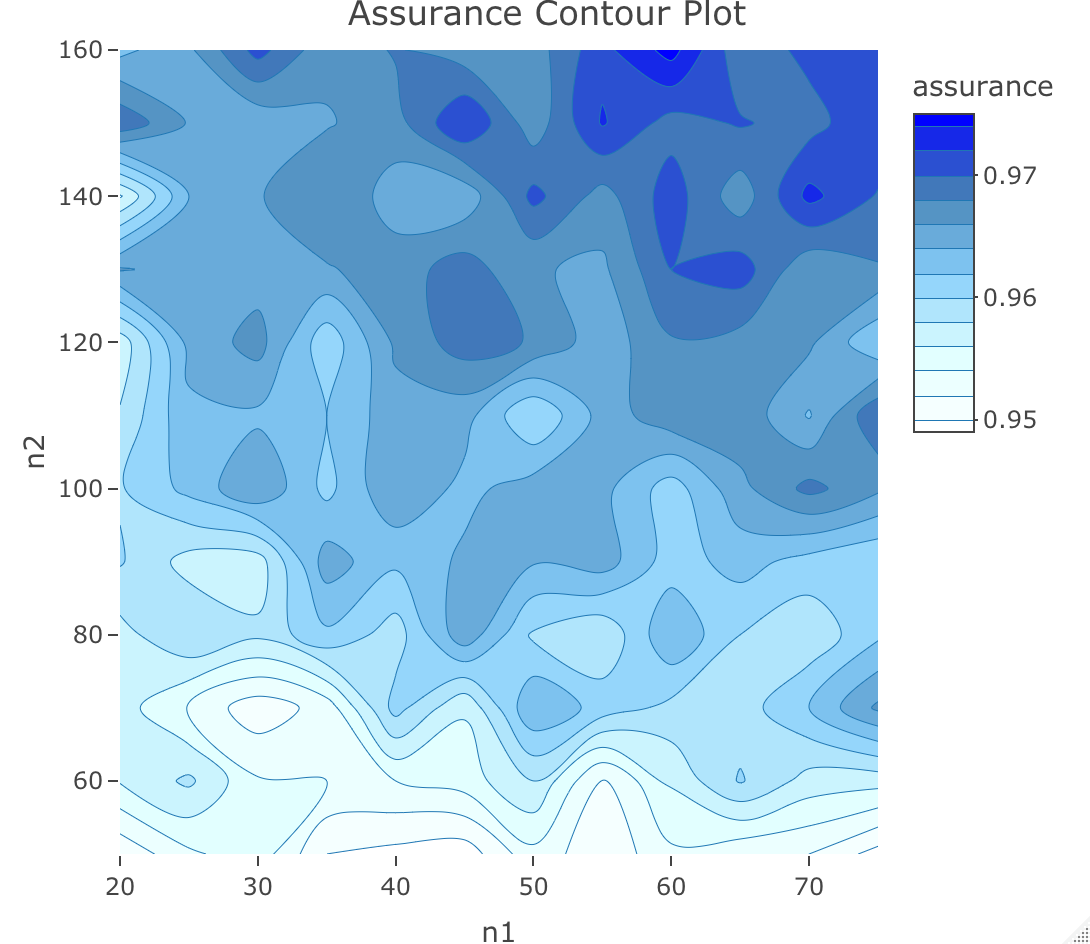}
\caption{\label{fig:ex5} Contour map of assurance values with
varying sample sizes \code{n1} and \code{n2}.}
\end{figure}

\subsubsection{Example 6: Cost-effectiveness Application}
We revisit the cost-effectiveness problem discussed in Example 2 of 
Section~\ref{sec:assur_knownvar}. In addition to providing
a 3-D graphical display of the assurance, 
this example also serves to demonstrate a setting where the 
\code{repeats} parameter becomes relevant. 

Recall from Example 2 that two distinct sets of efficacy and
cost measures are used to compare the cost-effectiveness 
of treatments 1 and 2. The efficacy and costs
are denoted by $\mu_i$ and $\gamma_i$ for 
$i = 1, 2$ treatments. Hence, the parameter we want to
estimate contains four elements tied to the unknown
efficacy and costs of treatments 1 and 2, i.e.
$\beta = (\mu_1, \gamma_1, \mu_2, \gamma_2)^{\top}$. 
It was previously assumed that the treatments contain an equal number of observations,
suggesting that the sample sizes across each of the four explanatory variables are also equal. Using \fct{bayes\_sim\_unbalanced} 
offers the added flexibility of constructing an unbalanced 
study design between treatments 1
and 2.  Since the two treatments each contain two components
to be measured, we use the \code{repeats} parameter to indicate
that we want two sets of sample sizes, \code{c(n1, n2)}, passed in, 
i.e. \code{c(n1, n2, n1, n2)}. It then becomes clear that our
study design consists of $n_1$ observations for the efficacy
and cost of treatment 1, and $n_2$ observations for those
of treatment 2.  Figure~\ref{fig:ex5} displays a contour plot
with a noticeable increasing trend of assurance values across 
larger sets of sample sizes. 

\begin{CodeChunk}
\begin{CodeInput}
R> library(bayesassurance)
R> n1 <- c(4, 5, 15, 25, 30, 100, 200)
R> n2 <- c(8, 10, 20, 40, 50, 200, 250)

R> mu_beta_d <- as.matrix(c(5, 6000, 6.5, 7200))
R> mu_beta_a <- as.matrix(rep(0, 4))
R> K = 20000 # threshold unit cost
R> C <- 0
R> u <- as.matrix(c(-K, 1, K, -1))
R> sigsq <- 4.04^2
R> Vbeta_a_inv <- matrix(rep(0, 16), nrow = 4, ncol = 4)
R> Vbeta_d <- (1 / sigsq) * matrix(c(4, 0, 3, 0, 0, 10^7, 0, 0, 
3, 0, 4, 0, 0, 0, 0, 10^7),nrow = 4, ncol = 4)

R> assur_out <- bayes_sim_unbalanced(n1 = n1, n2 = n2, repeats = 2, 
		u = as.matrix(c(-K, 1, K, -1)), C = 0, Xn = NULL, 
		Vbeta_d = Vbeta_d, Vbeta_a_inv = Vbeta_a_inv,
		Vn = NULL, sigsq = 4.04^2, 
		mu_beta_d = as.matrix(c(5, 6000, 6.5, 7200)), 
		mu_beta_a = as.matrix(rep(0, 4)), 
		alt = "greater", alpha = 0.05, mc_iter = 5000,  
		surface_plot = TRUE)
		
R> assur_out$assurance_table
R> assur_out$contourplot
\end{CodeInput}
\begin{CodeOutput}
   n1  n2 Assurance
1   4   8    0.1614
2   5  10    0.1724
3  15  20    0.3162
4  25  40    0.3942
5  30  50    0.4440
6 100 200    0.6184
7 200 250    0.7022
\end{CodeOutput}
\end{CodeChunk}

\begin{figure}[t!]
\centering
\includegraphics[width = 10 cm]{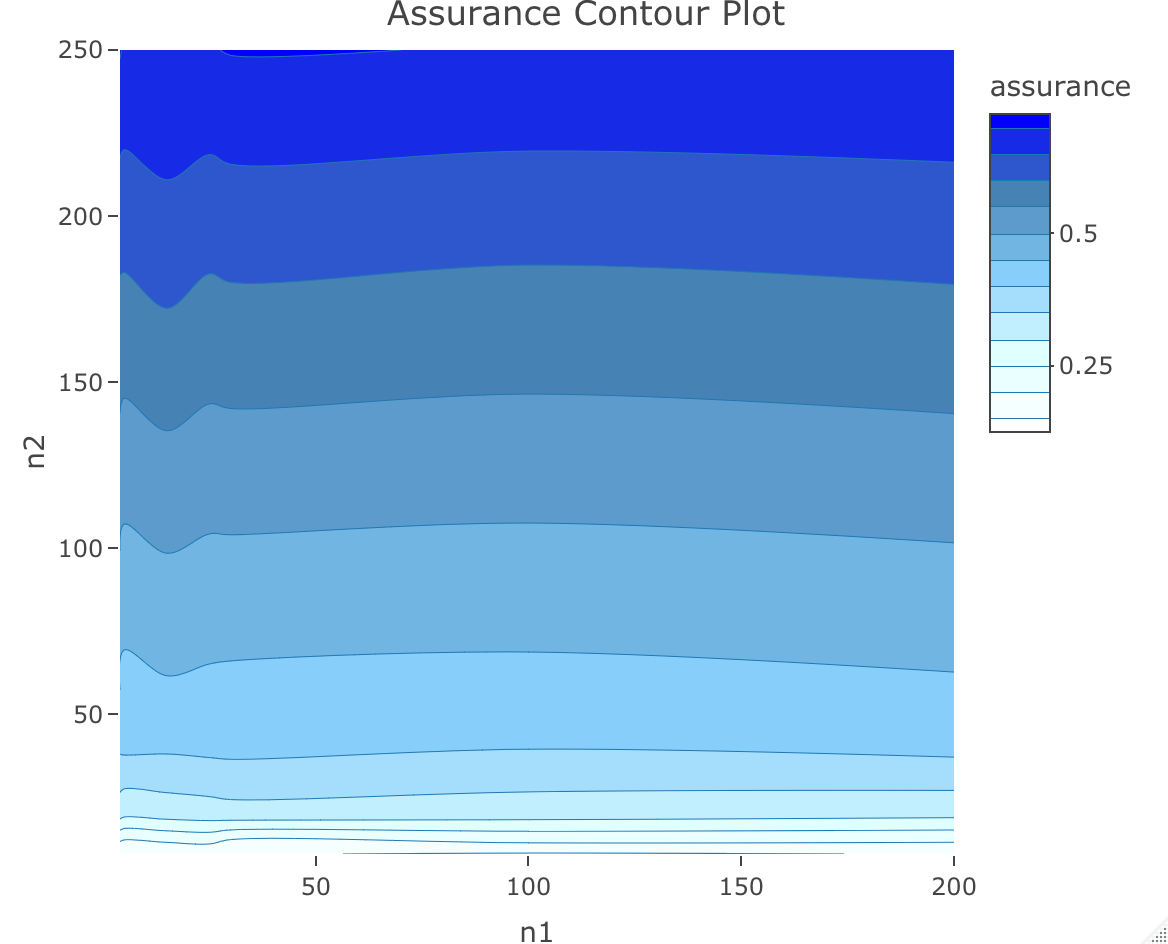}
\caption{\label{fig:ex5} Contour map of assurance values in
cost-effectiveness application.}
\end{figure}

\section{Bayesian Assurance Using Other Conditions}
\label{sec:othermethods}

\subsection{Bayesian Assurance with Precision-Based Conditions}

The \fct{bayes\_adcock} function determines the assurance under 
conditions discussed in~\cite{cj}. 
In the frequentist setting, if $X_i \sim N(\mu, \sigma^2)$ 
for $i = 1,..., n$ observations and variance $\sigma^2$ is known, 
the precision can be determined using 
$d = z_{1-\alpha/2}\frac{\sigma}{\sqrt{n}}$, 
where $z_{1-\alpha/2}$ is the critical value for the $100(1-\alpha/2)\%$ 
quartile of the standard normal distribution. 
In this setting, \textit{precision} refers to the allotted margin of
difference allotted between the true and estimated parameter values. 
Simple rearranging leads 
to the following expression for sample size,
\begin{equation}
\label{eq:precision_samplesize}
  n = z_{1-\alpha/2}^2\frac{\sigma^2}{d^2}\;.
\end{equation}
Given a random sample with mean $\bar{x}$, suppose the 
goal is to estimate population mean $\mu$.
To construct this problem in the Bayesian setting, 
suppose $x_1, \cdots, x_n$ 
is a random sample from $N(\mu, \sigma^2)$
and the sample mean is distributed as $\bar{x}|\mu, \sigma^2 \sim 
N(\mu, \sigma^2/n)$.
The analysis objective is to observe that the absolute 
difference between $\bar{x}$ and $\mu$ falls within a margin of 
error no greater than $d$. Given data $\bar{x}$ and a pre-specified 
confidence level $\alpha$, the assurance can be formally expressed as
\begin{equation}
\label{eq:assur_adcock}
\delta = P_{\bar{x}}\{\bar{x}: P(|\bar{x} - \mu| \leq d) \geq 1-\alpha\}\;.
\end{equation}
We assign $\mu \sim N(\mu_{\beta}^{(a)}, \sigma^2/n_a)$ as the 
analysis stage prior, where $n_a$ quantifies the amount of prior 
information we have for $\mu$.
The posterior of $\mu$ is obtained by taking the product of the prior 
and likelihood, resulting in
\begin{equation}
\nonumber
  N\Bigg(\bar{x} {\left| \mu, \frac{\sigma^2}{n}\right.}\Bigg) \times
  N\Bigg(\mu {\left | \mu_{\beta}^{(a)}, \frac{\sigma^2}{n_a}\right.}\Bigg)
  = N\Bigg(\mu {\left | \lambda, \frac{\sigma^2}{n_a+n}\right.}\Bigg),
 \end{equation}
where $\lambda = \frac{n\bar{x}+n_a\mu_{\beta}^{(a)}}{n_a+n}$.
We can further evaluate the condition using parameters from the posterior of $\mu$ to obtain a more explicit version of the analysis stage objective. Starting from the condition expressed in 
Equation~\eqref{eq:assur_adcock}, where
$P(|\bar{x} - \mu| \leq d) = P(\bar{x} - d \leq \mu \leq \bar{x} + d)$, 
we can standardize all components of the inequality using 
the posterior distribution of $\mu$, eventually leading to
\begin{equation}
\nonumber
  P(|\bar{x} - \mu| \leq d) =
  \Phi\left[\frac{\sqrt{n_a + n}}{\sigma}
  (\bar{x} + d - \lambda)\right] - \Phi\left[\frac{\sqrt{n_a+n}}{\sigma}(\bar{x}
  - d - \lambda)\right],
\end{equation}
where $\Phi$ denotes the cumulative density function of the 
standard normal distribution.

We now assign a separate design stage prior on $\mu$ such that 
$\mu \sim N(\mu_{\beta}^{(d)}, \sigma^2/n_d)$, where $n_d$ quantifies
our degree of belief towards the population from which the sample 
will be drawn. Given that $\bar{x}|\mu, \sigma^2 \sim N(\mu, \sigma^2/n)$, 
the marginal distribution of $\bar{x}$ can be computed using straightforward 
substitution based on $\bar{x} = \mu + \epsilon;$ $\epsilon \sim N(0, \sigma^2/n)$ 
and $\mu = \mu_{\beta}^{(d)} + \omega;$ $\omega \sim N(0, \sigma^2/n_d)$.  
Substituting $\mu$ into the expression for $\bar{x}$ gives us
$\bar{x}=\mu_{\beta}^{(d)} + (\omega + \epsilon);$
$(\omega + \epsilon) \sim N\Big(0, \frac{\sigma^2(n_d+n)}{n_dn}\Big)
= N(0, \sigma^2/p)$, where $1/p = 1/n_d + 1/n$. The marginal
of $\bar{x}$ is therefore $N(\bar{x}|\mu_{\beta}^{(d)}, \sigma^2/p)$, 
denoting the distribution in which we will 
iteratively generate samples from to check if they satisfy the condition
outlined in the analysis stage and the assurance is computed as
\begin{equation}
  \label{eq:adcock_region_simplified}
 \hat{\delta}(n) = \frac{1}{J}\sum_{j=1}^J\mathbb{I}\left(\left\{ \bar{x}_{(j)}:
  \Phi\left[\frac{\sqrt{n_a + n}}{\sigma}
  (\bar{x} + d - \lambda)\right] - \Phi\left[\frac{\sqrt{n_a+n}}{\sigma}(\bar{x}
  - d - \lambda)\right] \geq 1-\alpha \right\}\right).
\end{equation}

The \fct{bayes\_adcock} function takes in the following 
set of parameters:
\begin{enumerate}
\item{\code{n}: sample size (either vector or scalar)}
\item{\code{d}: precision level expressed 
in analysis objective, $|\bar{x} - \mu| \leq d$}
\item{\code{mu_beta_a}: analysis stage mean}
\item{\code{mu_beta_d}: design stage mean}
\item{\code{n_a}: sample size at analysis stage; quantifies the 
amount of prior information we have for parameter $\mu$.}
\item{\code{n_d}: sample size at design stage; 
quantifies the amount of prior 
information we have for where the data is being generated from.}
\item{\code{sig_sq}: known variance $\sigma^2$.}
\item{\code{alpha}: significance level}
\item{\code{mc_iter}: number of MC samples evaluated 
under the analysis objective. }
\end{enumerate}

\subsubsection{Example 7: Precision-Based Condition}
\begin{figure}[t!]
\centering
\includegraphics[width = 8 cm]{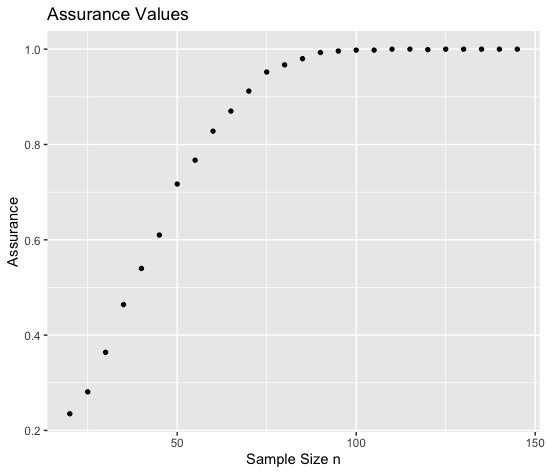}
\caption{\label{fig:ex7} Assurance points simulated
from \fct{bayes\_adcock} in Example 7.}
\end{figure}

As a preliminary example, consider passing in the following 
set of arbitrary inputs into \fct{bayes\_adcock}, with the outputs
saved as \code{out}. Since we are assigning a vector of 
values to \code{n}, we should therefore expect a vector
of assurance values being returned.
The first six rows of the assurance table
is shown below and the plotted assurance points are displayed
in Figure~\ref{fig:ex7}.

\begin{CodeChunk}
\begin{CodeInput}
R> library(bayesassurance)
R> n <- seq(20, 145, 5)
R> out <- bayes_adcock(n = n, d = 0.20, mu_beta_a = 0.64, mu_beta_d = 0.9, 
                                    n_a = 20, n_d = 10, sig_sq = 0.265, 
                                    alpha = 0.05, mc_iter = 10000)                                    
R> head(out$assurance_table)
R> out$assurance_plot
\end{CodeInput}      
\begin{CodeOutput}
   n Assurance
1 20     0.235
2 25     0.281
3 30     0.364
4 35     0.464
5 40     0.540
6 45     0.610
\end{CodeOutput}                              
\end{CodeChunk}   

\subsubsection{Example 8: Overlap Between Classical and Bayesian Setting}
To demonstrate how the frequentist setting is perceived as a special 
case of the Bayesian setting under the 
precision-based conditions discussed above,
we refer to the sample size formula
in Equation~\eqref{eq:precision_samplesize}
and apply rudimentary rearrangements to obtain
$2\Phi\Big[\frac{\sqrt{n}}{\sigma}d\Big]-1 \geq 1-\alpha$,
which we denote as the frequentist condition. 
\begin{figure}[t!]
\centering
\includegraphics[width = 12 cm]{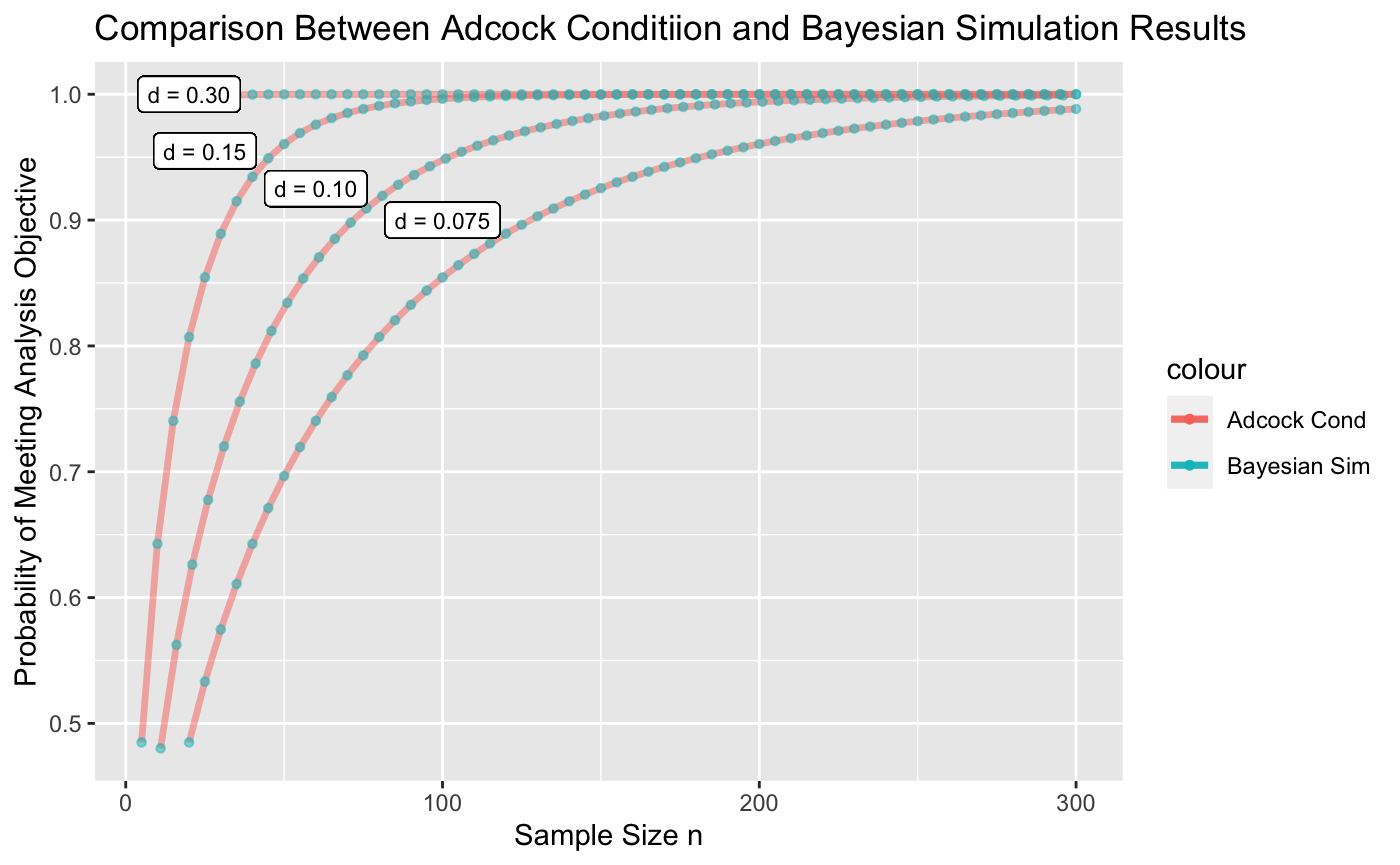}
\caption{\label{fig:ex8} Left-hand-side expressions of
frequentist and Bayesian cases under precision-based 
conditions for varying precision levels, $d$.}
\end{figure}
Comparing this to the assurance formula
expressed in Equation~\eqref{eq:adcock_region_simplified}, where
\begin{equation}
\nonumber
  \delta = \left\{ \bar{x}: \Phi\left[\frac{\sqrt{n_a + n}}{\sigma}
  (\bar{x} + d - \lambda)\right] - \Phi\left[\frac{\sqrt{n_a+n}}{\sigma}
  (\bar{x} - d - \lambda)\right] \geq 1-\alpha \right\},
\end{equation}
we notice that both expressions have $1-\alpha$ isolated 
on the right hand side. 
Setting $n_a = 0$ in the assurance formula leads to 
\begin{align*}
&\delta = \left\{ \bar{x}: \Phi\left[\frac{\sqrt{n_a + n}}{\sigma}
  (\bar{x} + d - \lambda)\right] - \Phi\left[\frac{\sqrt{n_a+n}}{\sigma}(\bar{x}
  - d - \lambda)\right] \geq 1-\alpha \right\} \\
  &\xrightarrow[]{n_a = 0} \left\{ \bar{x}: \Phi\left[\frac{\sqrt{n}}
  {\sigma}d\right] - \Phi \left[-\frac{\sqrt{n}}{\sigma}d\right] 
  \geq 1 - \alpha \right\} = \left\{
  \bar{x}: 2\Phi \left[\frac{\sqrt{n}}{\sigma}d\right] - 1 \geq 1 - \alpha \right\}.
\end{align*}
The resulting term on the left hand side is now equivalent 
to that of the frequentist condition, suggesting that if we let $\theta$ take on a weak analysis prior, i.e. $n_a = 0$,our analysis objective
simplifies to an expression that directly aligns with the frequentist
formulation of sample size.   

A visual portrayal of this special case is shown in 
Figure~\ref{fig:ex8}, which plots the left-hand-side 
expressions being compared to $1-\alpha$ in
both settings. Note that these expressions do not correspond to the 
actual assurance values, but rather, the probability of meeting the 
specified analysis objective within a single iteration. The assurance
takes the average of the indicator-based outcomes 
across all iterations, as shown in Equation~\eqref{eq:adcock_region_simplified}.  
Red curves correspond to values obtained from 
the frequentist expression, $2\Phi\Big[\frac{\sqrt{n}}{\sigma}d\Big]-1$,
while blue points denote results determined from 
the Bayesian expression, 
$\Phi\left[\frac{\sqrt{n_a + n}}{\sigma} (\bar{x} + d - \lambda)\right] -
\Phi\left[\frac{\sqrt{n_a+n}}{\sigma}(\bar{x} - d - \lambda)\right]$, 
which uses samples generated from the marginal of $\bar{x}$ and 
assumes a weak analysis prior.  This is considered for multiple
precision levels, $d$.


\subsection{Bayesian Assurance Using Credible Interval Based Conditions}\label{sec:credible_interval_method}
The \fct{bayes\_sim\_betabin} function determines assurance
in the context of estimating differences between two proportions.  
The analysis objective in the Bayesian setting relies on credible 
interval based conditions discussed in \cite{pham} that 
outlines steps for determining exact sample sizes
needed in this particular setting.

Starting with the classical setting, let $p_i, i = 1, 2$ denote 
two independent proportions pulled from distinct samples of size
$n$. Suppose we want to execute the hypothesis test, 
$H_0: p_1 - p_2 = 0$ vs. $H_a: p_1 - p_2 \neq 0$.
The outcome of interest involves involves detecting 
if a significant difference exists between the two proportions. 
One way to check is to see if $0$ is contained
within the confidence bands of the difference
in proportions given by 
\begin{equation}
\label{eq:prop_classical_ci}
0 \in \displaystyle (\hat{p_1} - \hat{p_2}) \pm z_{1-\alpha/2}(SE(\hat{p_1})^2 + SE(\hat{p_2})^2)^{1/2},
\end{equation}
where $z_{1-\alpha/2}$ denotes the critical region, and $SE(\hat{p_i})$
denotes the standard error of $p_i, i = 1, 2$, calculated as 
$SE(\hat{p_i}) = \sqrt{\frac{(\hat{p_i})(1-\hat{p_i})}{n}}$. 
An interval without $0$ included in the confidence 
interval suggests there exists a significant difference 
between the two proportions.

The Bayesian setting uses posterior credible intervals 
as an analog to the frequentist confidence interval approach.
Two distinct priors are assigned to $p_1$ and $p_2$ 
such that $p_i \sim \text{Beta}(\alpha_i, \beta_i)$ for $i = 1,2$.
Let $X$ be a random variable taking on values $x = 0, 1,..., n$ 
denoting the number of favorable
outcomes out of $n$ trials. The proportion of favorable outcomes is
therefore $p = x/n$. Suppose a Beta prior is assigned to $p$
such that  $p \sim \text{Beta}(\alpha, \beta)$. The prior mean and 
variance are respectively $\mu_{prior} = \frac{\alpha}{\alpha + \beta}$ 
and $\sigma^2_{prior} = \frac{\alpha\beta}{(\alpha + \beta)^2(\alpha +
\beta + 1)}$. Conveniently, given that $p$ is assigned a Beta prior, 
the posterior of $p$ also takes on a Beta distribution with 
mean and variance
\begin{equation}
\label{eq:beta_post_parameters}
    \mu_{posterior} = \frac{\alpha + x}{\alpha + \beta + n}; \quad
    \sigma^2_{posterior} = \frac{(\alpha + x)(\beta + n - x)}{(\alpha + \beta + n)^2(\alpha + \beta + n + 1)}.
\end{equation}


In the analysis stage, we assign priors for 
$p_1$ and $p_2$ such that
$p_i \sim \text{Beta}(\alpha_i, \beta_i), i = 1, 2$. Letting $p_{d} = 
p_1 - p_2$ and having 
$p_{\text{post}}$ and $\text{var}(p)_{\text{post}}$
respectively denote the posterior mean and variance of $p_{d}$,
it is straightforward to deduce from 
Equation~\eqref{eq:beta_post_parameters} that $p_{\text{post}} = \frac{\alpha_1 + x_1}{\alpha_1 + \beta_1 + n_1} -
\frac{\alpha_2 + x_2}{\alpha_2 + \beta_2 + n_2}$ and $\text{var}(p)_{\text{post}} =
\frac{(\alpha_1 + x_1)(\beta_1 + n_1 - x_1)}{(\alpha_1 + \beta_1 + n_1)^2(\alpha_1 + \beta_1 + n_1 + 1)} +
\frac{(\alpha_2 + x_2)(\beta_2 + n_2 - x_2)}{(\alpha_2 + \beta_2 + n_2)^2(\alpha_2 + \beta_2 + n_2 + 1)}.$
The resulting $100(1-\alpha)\%$ credible interval therefore 
equates to $p_{\text{post}} \pm z_{1-\alpha/2}
\sqrt{\text{var}(p)_{\text{post}}}$, which, similar
to the frequentist setting, is used to check if 
$0$ is contained within the credible interval bands as 
part of our inference procedure. Our analysis objective becomes 
\begin{equation}
\label{eq: region_cond}
R(x_1, x_2) = \left\{x_1, x_2: 0 \not\in
\left(p_{\text{post}} - z_{1-\alpha/2} \sqrt{\text{var}(p)_{\text{post}}}, \quad p_{\text{post}} +
z_{1-\alpha/2} \sqrt{\text{var}(p)_{\text{post}}}\right)\right\},
\end{equation}
where $x_1$ and $x_2$ are the counts observed 
in the two samples generated in the design stage. 


In the design stage, counts $x_1$ and $x_2$ are observed 
from samples of size $n_1$ and $n_2$
based on given probabilities, $p_1$ and $p_2$, provided in the
analysis stage. Once $p_1$ and $p_2$ are assigned, $x_1$ 
and $x_2$ are randomly generated from
their corresponding binomial distributions, where 
$x_i \sim \text{Bin}(n_i, p_i), i = 1,2$. The
posterior credible intervals are subsequently computed to undergo assessment in
the analysis stage. These steps are repeated iteratively starting from the generated values of 
$x_1$ and $x_2$. The proportion of samples that meet the criterion in
Equation~\eqref{eq: region_cond} equates to the assurance.
It follows that the corresponding assurance for assessing a 
significant difference in proportions is given by 
\[
\delta = P_{x_1, x_2} \left\{x_1, x_2: P\left(0 \not\in
\left(p_{\text{post}} - z_{1-\alpha/2} \sqrt{\text{var}(p)_{\text{post}}},
\quad p_{\text{post}} +
z_{1-\alpha/2} \sqrt{\text{var}(p)_{\text{post}}}\right)\right) \geq 1-\alpha \right\}.
\]

The \fct{bayes\_sim\_betabin} function takes
the following set of inputs: 
\begin{enumerate}
\item{\code{n1}: sample size of first group (scalar)}
\item{\code{n2}: sample size of second group (scalar)}
\item{\code{p1}: proportion of successes in first group; set 
to \code{NULL} by default}
\item{\code{p2}: proportion of successes in second group; set 
to \code{NULL} by default}
\item{\code{alpha_1, beta_1}: shape parameters for the 
distribution of \code{p1} such that $p1 \sim \text{Beta}(\alpha_1, \beta_1)$; used when \code{p1} is not
provided or known}
\item{\code{alpha_2, beta_2}: shape parameters for the 
distribution of \code{p2} such that $p2 \sim \text{Beta}(\alpha_12 \beta_2)$; used when \code{p2} is not
provided or known}
\item{\code{sig_level}: significance level}
\item{\code{alt}: a character string specifying the alternative hypothesis when comparing \code{p1}
and \code{p2}. The options are \code{"two.sided"} (default), 
\code{"greater"} and  \code{"less"}.}
\item{\code{mc_iter}: number of MC samples evaluated 
under the analysis objective; set to 5000 by default.}
\end{enumerate}

\subsubsection{Example 9: Simple Case}
\begin{figure}[t!]
\centering
\includegraphics[width = 7 cm]{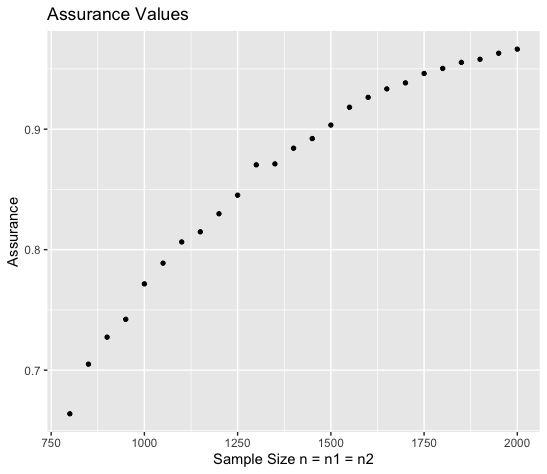}
\caption{\label{fig:ex9} Simulated assurance points corresponding
to Example 9 using credible interval based conditions.}
\end{figure}

In the following example, let \code{n1 = n2} for simplicity and choose 
arbitrary values for the remaining parameters.  
Results are saved as \code{out}, containing 
corresponding plots and tables.  The first
six rows of the assurance table is shown below
and Figure~\ref{fig:ex9} plots estimated
assurances for varying sample sizes. 

\begin{CodeChunk}
\begin{CodeInput}
R> n <- seq(800, 2000, 50)

R> out <- bayes_sim_betabin(n1 = n, n2 = n, p1 = 0.25, p2 = 0.2, 
                            alpha_1 = 0.5, beta_1 = 0.5, alpha_2 = 0.5, 
                            beta_2 = 0.5, sig_level = 0.05, alt = "two.sided")

R> head(out$assurance_table)
R> out$assurance_plot      
                                                                   
\end{CodeInput}
\begin{CodeOutput}
     n1   n2 Assurance
1   800  800    0.6638
2   850  850    0.7050
3   900  900    0.7274
4   950  950    0.7422
5  1000 1000    0.7716
6  1050 1050    0.7888
\end{CodeOutput}
\end{CodeChunk}         

The next two examples demonstrate cases when 
the Bayesian and frequentist settings perform similarly
in behavior. 
                      
\subsubsection{Example 10: Special Case 1}    

If $p_1$ and $p_2$ are known, there is no need generate 
proportions from the Beta distribution as outlined in the 
beginning of Section~\ref{sec:credible_interval_method}. 
We still, however, need to assign shape parameters 
to determine the posterior credible interval bands.  
In this example, the shape parameters of the two groups are
assigned $\alpha_1 = \beta_1 = \alpha_2 = \beta_2 = 0.5$ in
\fct{bayes\_sim\_betabin}. These 
specifications are often referred to as Haldane's priors, or flat 
priors, of the Beta distribution. 

Recall the sample size formula for assessing
differences in proportions in the frequentist setting,
\[
  n = \frac{(z_{1-\alpha/2} + z_{\beta})^2(p_1(1-p_1) + p_2(1-p_2))}{(p_1 - p_2)^2},
  \]
where $n = n_1 = n_2$. Simple rearrangements and also noting that
$-(z_{1-\alpha/2} + z_{\beta}) = z_{1-\beta} - z_{1-\alpha/2}$ lead to
\begin{multline}
\label{eq:prop_power}
\frac{\sqrt{n}(p_1 - p_2)}{\sqrt{p_1(1-p_1) + p_2(1-p_2)}} + z_{1-\alpha/2} = z_{1-\beta}
\implies
\text{Power} = 1 - \beta
= \\ \Phi\left(\frac{\sqrt{n}(p_1 - p_2)}{\sqrt{p_1(1-p_1) + p_2(1-p_2)}} + z_{1-\alpha/2}\right).
\end{multline}

Similar to Bayesian assurance, power 
is interpreted as the probability of correctly detecting a 
specified difference between the two
proportions, $p_1$ and $p_2$, in correspondence
to sample size $n$. We translate this relationship in to a 
simple function that takes a set of fixed parameters 
and returns the power based on Equation~\eqref{eq:prop_power}: 
\begin{CodeChunk}
\begin{CodeInput}
R> propdiffCI_classic <- function(n, p1, p2, sig_level){
  p <- p1 - p2
  power <- pnorm(sqrt(n / ((p1*(1-p1)+p2*(1-p2)) / (p)^2)) - 
  qnorm(1-sig_level/2))
  return(power)
}
\end{CodeInput}
\end{CodeChunk}

The next code segment determines power and assurance 
for a fixed set of inputs across a vector of sample sizes. 

\begin{CodeChunk}
\begin{CodeInput}
R> n <- seq(40, 1000, 10)

## Power Computation
R> power_vals <- propdiffCI_classic(n=n, p1 = 0.25, p2 = 0.2, 
sig_level = 0.05)
R> power_table <- as.data.frame(cbind(n, power_vals))

## Assurance Computation
R> assurance_out <- bayes_sim_betabin(n1 = n, n2 = n, p1 = 0.25, 
              p2 = 0.2, alpha_1 = 0.5, beta_1 = 0.5, alpha_2 = 0.5, 
              beta_2 = 0.5, sig_level = 0.05, alt = "two.sided")
R> assurance_table <- assurance_out$assurance_table
R> colnames(assurance_table) <- c("n1", "n2", "Assurance")
\end{CodeInput}
\end{CodeChunk}

\begin{figure}[t!]
\centering
\includegraphics[width = 12 cm]{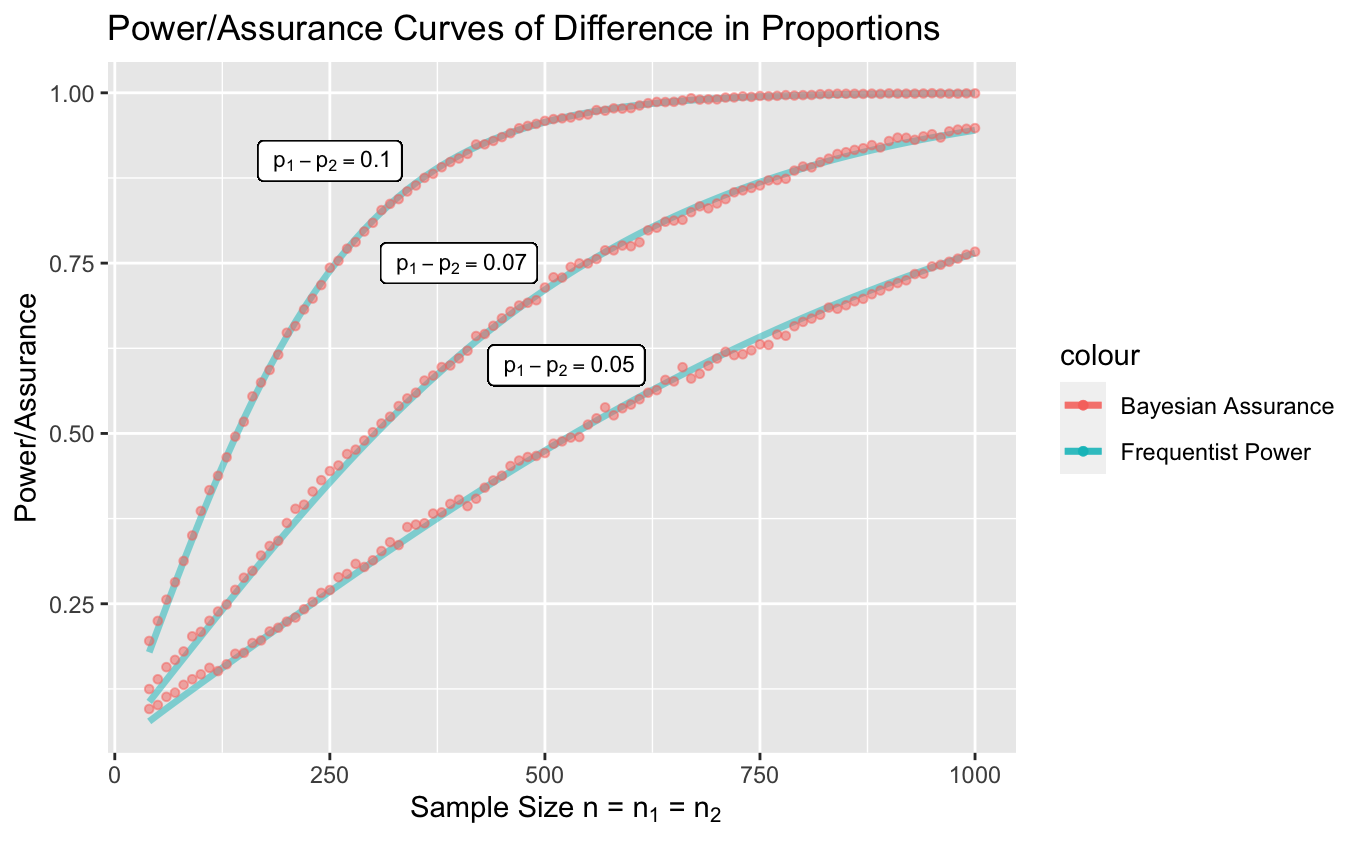}
\caption{\label{fig:ex10} Power curves and simulated 
assurance points corresponding to the case when $p_1$ 
and $p_2$ are known.}
\end{figure}

We repeat these steps for different values of $p_1$ and $p_2$ while
keeping the same flat priors. As shown in Figure~\ref{fig:ex10},
we do this for assumed differences of 0.1, 0.07, and 0.05. Since it
is more difficult to detect smaller differences, higher sample sizes are
needed to ensure that we are appropriately identifying fulfillment 
of the analysis objective expressed in Equation~\ref{eq: region_cond}. 

\subsubsection{Example 11: Special Case 2}   

The next example demonstrates overlapping behaviors of the
frequentist and Bayesian settings when $p_1$ and $p_2$ are 
unknown. In an ideal situation, we could determine 
suitable parameters for $\alpha_i$, $\beta_i, i = 1, 2$ 
to use as our Beta priors that would enable 
demonstration of convergence towards the
frequentist setting. However, a key relationship to recognize
is that the Beta distribution is a conjugate prior of the Binomial
distribution. There is a subtle advantage offered given
that the Bayesian credible interval bands are based upon
posterior parameters of the Beta distribution 
(refer to Equation~\eqref{eq: region_cond}) and the
frequentist confidence interval bounds are based upon
the Binomial distribution (refer to Equation~\eqref{eq:prop_classical_ci}). 
Because of the conjugate relationship
held by the Beta and Binomial distributions, we are 
assigning priors to parameters in the Bayesian setting
that the Binomial density in the frequentist setting
is conditioned upon. Since the normal distribution
can be used to approximate binomial distributions for large
sample sizes given that the Beta distribution is approximately
normal when its parameters $\alpha$ and $\beta$ are set to 
be equal and large, we manually choose such values to 
illustrate these parallels. 
First, we modify the frequentist power function 
from before to now include shape parameters: 


\begin{CodeChunk}
\begin{CodeInput}
R> propdiffCI_classic <- function(n, p1, p2, alpha_1, beta_1, alpha_2,
beta_2, sig_level){
  set.seed(1)
  if(is.null(p1) == TRUE & is.null(p2) == TRUE){
    p1 <- rbeta(n=1, alpha_1, beta_1)
    p2 <- rbeta(n=1, alpha_2, beta_2)
  }else if(is.null(p1) == TRUE & is.null(p2) == FALSE){
    p1 <- rbeta(n=1, alpha_1, beta_1)
  }else if(is.null(p1) == FALSE & is.null(p2) == TRUE){
    p2 <- rbeta(n=1, alpha_2, beta_2)
  }
  p <- p1 - p2

  power <- pnorm(sqrt(n / ((p1*(1-p1)+p2*(1-p2)) / (p)^2)) - 
  qnorm(1-sig_level/2))
  return(power)
}
\end{CodeInput}
\end{CodeChunk}

The following code sets parameters 
$p_1$ and $p_2$ to \code{NULL} and assigns equal 
sets of arbitrarily selected shape parameters 
to pass in to the assurance function, where 
$\alpha_1 = \beta_1 = 2$ and $\alpha_2 = \beta_2 = 6$. 

\begin{CodeChunk}
\begin{CodeInput}
################################################
# alpha1 = 2, beta1 = 2, alpha2 = 6, beta2 = 6 #
################################################

R> power_vals <- propdiffCI_classic(n=n, p1 = NULL, p2 = NULL, 2, 2, 6, 6, 0.05)
R> df1 <- as.data.frame(cbind(n, power_vals))

R> assurance_vals <- bayes_sim_betabin(n1 = n, n2 = n, 
              p1 = NULL, p2 = NULL, alpha_1 = 2, beta_1 = 2, alpha_2 = 6, 
              beta_2 = 6, sig_level = 0.05, alt = "two.sided")
R> df2 <- assurance_vals$assurance_table
\end{CodeInput}
\end{CodeChunk}

\begin{figure}[t!]
\centering
\includegraphics[width = 12 cm]{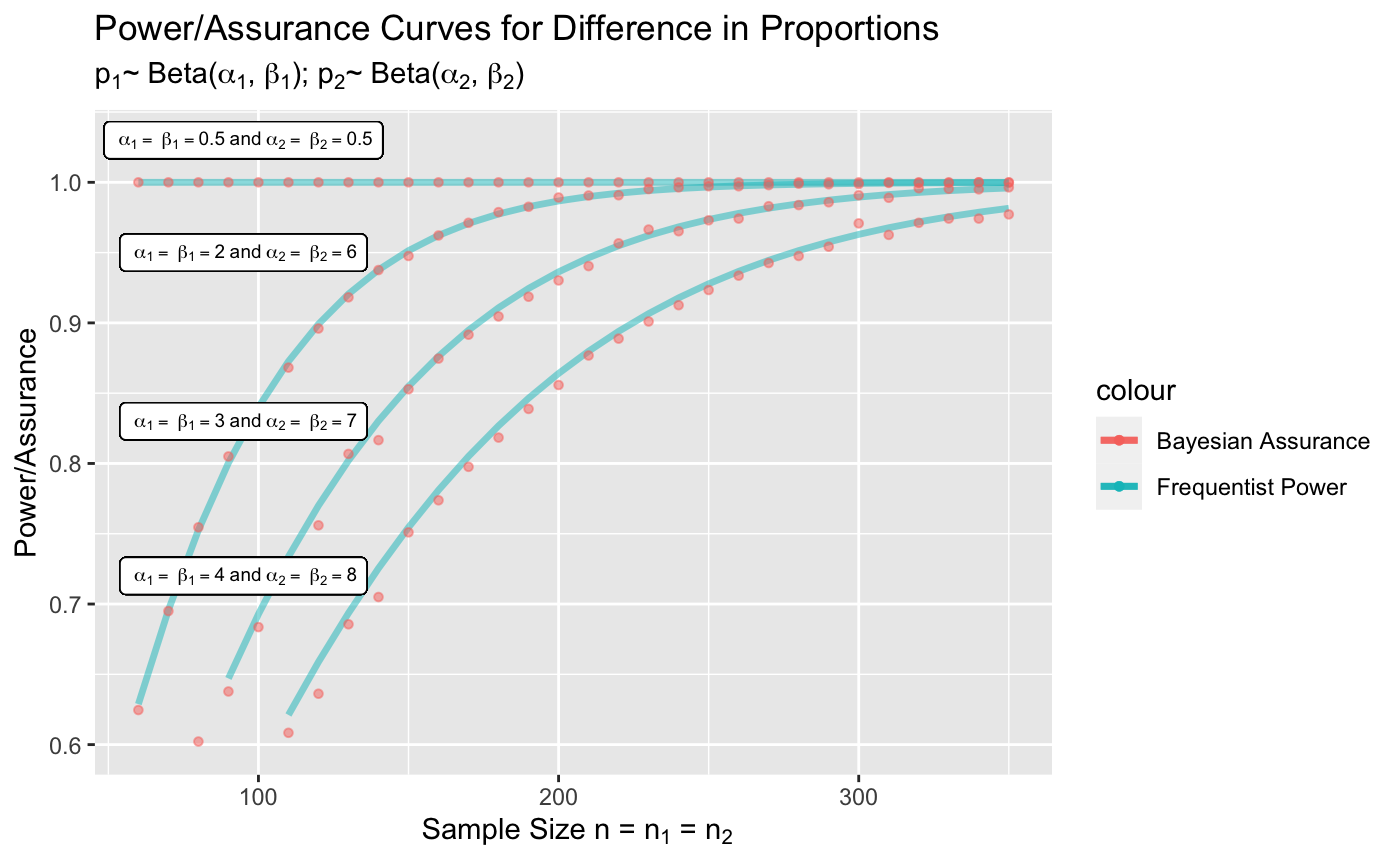}
\caption{\label{fig:ex11} Power curves and simulated 
assurance points corresponding to the case when $p_1$ 
and $p_2$ are unknown. }
\end{figure}  

We repeat these steps for multiple sets of 
shape parameters that are also set equal to each other.
The results are gathered into separate tables reporting
sample sizes and corresponding power and assurance values
for all the different cases.These tables are used to construct 
Figure~\ref{fig:ex11}, which simultaneously plots 
estimated assurance values, denoted by
red points,laying along the respective power curves, 
marked in blue. 

Notice that the power and assurance values 
approximately overlap each other in each of the four cases, 
which is particularly true for the case when Haldane priors are used. 
This again relies on the notion that the normal distribution can be 
used to approximate binomial distributions for large
sample sizes given that the Beta distribution is approximately 
normal when its parameters $\alpha$ and $\beta$ are set to be equal.

\section{Bayesian Goal Function}\label{sec: goalfunc}

Similar to the Bayesian assurance, goal functions seek to determine 
the minimum sample size needed to meet certain objectives. 
Discussed in~\cite{inoue}, the goal function aggregates
well-defined conditions that are dependent on sample size
and other user-specified inputs relevant to the study. It is framed under a 
utility approach that enables the decision maker to choose
an option that maximizes the expected value of the
function. The goal function 
specifies an overarching condition that the study aims 
to observe, and in turn, assists users in making
well-informed decisions regarding their analysis. 
Specification of goal functions comes in a variety of forms,
including but not limited to, statistical power, information, 
confidence interval widths and rate of classification error
\citep{inoue}.The goal functions defined in our package
are characterized by the use of the rate of correct 
classification, which we denote as $r^{*}$. 

Optimizing the value of $r^{*}$ occurs from either correctly 
rejecting the null hypothesis or correctly accepting the 
null hypothesis. Table~\ref{tab:utility_table} specifies 
the utility associated with all possible events. 
Specifically, a utility of $K$ is associated with correctly failing to 
reject $H_0$, a loss of 1 is associated with correctly
rejecting $H_0$, and incorrect decisions do not experience any 
benefit. A general expression of the goal function is given as 
\begin{equation} \label{eq:rateofclass}
G_B(n, \textbf{v}) = KP(H_0) P(\text{fail to reject } H_0 | \text{ } H_0 \text{ is true}) + P(H_a) P(\text{reject } H_0 | \text{ } H_a \text{ is true}),
\end{equation}
where $\textbf{v}$ denotes a vector of user-specified inputs 
in addition to sample size $n$. The value
of $G_B(n, \textbf{v})$ obtained 
once the inputs are processed is the overall $r^{*}$ value. 
\begin{table}[t!]
\centering
\begin{tabular}{ccc}
\hline
& $H_0$ is True  & $H_a$ is True \\ \hline
Do not Reject $H_0$ & $K$ & 0 \\ 
Reject $H_0$ & 0 & 1
 \\ \hline
\end{tabular}
\caption{\label{tab:utility_table} Utility weights assigned for 
different pairs of the true result (columns) and the 
decision (rows).}
\end{table}

Consider the linear hypothesis test $H_0: u^{\top}\beta = c_0$ 
vs.  $H_a: u^{\top}\beta = c_1$ under the general linear model 
$y = X\beta + \epsilon$, where $y$ is $n \times 1$, $X$ is 
$n \times p$, $\beta$ is $p \times 1$, $u$ is $p \times 1$, and 
$\epsilon \sim N(0, \sigma^2 I_n)$, implying that $c_0$
and $c_1$ are scalars. Let us assume that $u^{\top}\beta$ is
estimable. Then, by the fundamental principle of 
estimable functions, there exists a linear unbiased estimate
$b^{\top}y$ such that $E(b^{\top}y) = u^{\top}\beta$. This
suggests that $u$ belongs in the column space of $X^{\top}$ since
\[
E(b^{\top}y) = b^{\top}X\beta = u^{\top}\beta \implies
b^{\top}X = u^{\top} \implies u = X^{\top}b \implies
u \in C(X^{\top}).
\]
Hence, $u$ can be expressed as $u = X^{\top}z$ for some
$z \in \mathbb{R}^n$. 
Letting $\tilde{y} = z^{\top}y$, 
simple linear transformation leads to $\tilde{y} | H_0 \sim 
N(c_0, \sigma^2z^{\top}z)$ and $\tilde{y} | H_a \sim 
N(c_1, \sigma^2z^{\top}z)$. 
Under this framework, we 
treat Equation~\eqref{eq:rateofclass} as a template
and define specific thresholds that enable 
the function to objectively determine the rate of 
correct classification in the Bayesian setting.  
We assign a prior on $u^{\top}\beta$ such that 
$P(H_0) = 1 - P(H_a) = \pi$ and
assume that the null hypothesis is not rejected if
the posterior probability of $H_0$ is at least $1/(1+K)$, where
$K$ is the amount of utility associated with $H_0$ being 
correctly accepted.  Starting from the expression 
$P(H_0 | \tilde{y}) \geq \frac{1}{1+K}$, we apply
fundamental Bayesian principles on the left hand side 
of the inequality to obtain
\[
\frac{P(\tilde{y}|H_0)P(H_0)}{P(\tilde{y})} \geq \frac{1}{1+K}.
\]
Substituting appropriate densities and assigned values results in
\[
\frac{N(\tilde{y} | c_0, \sigma^2z^{\top}z)(\pi)}{N(\tilde{y} | c_0, \sigma^2z^{\top}z)(\pi) + N(\tilde{y} | c_1, \sigma^2z^{\top}z)(1-\pi)}
\geq \frac{1}{1+K}.
\]
Fundamental algebra leads to the following criteria, in which
$H_0$ is not rejected if 
\begin{equation}\nonumber
\tilde{y} \leq \frac{\sigma^2z^{\top}z}{\delta}\text{ln}\left(
\frac{K\pi}{1-\pi}\right) + \frac{c_1 + c_0}{2},
\end{equation}
where $\delta = c_1 - c_0$.
This condition can be expressed in a more cohesive
way through standardization. 
Given that $\tilde{y} \sim N(u^{\top}\beta, \sigma^2z^{\top}z)$,
it follows that 
\[
\frac{\tilde{y} - u^{\top}\beta}{\sigma\sqrt{z^{\top}z}} \sim
N(0, 1).
\]
Hence, the probability of correctly accepting $H_0$
is given by 
\begin{equation}\nonumber
P(\text{fail to reject } H_0 | \text{ } H_0 \text{ is true})=\Phi\left[\frac{\sigma\sqrt{z^{\top}z}}{\delta}
\text{ln}\left(\frac{K\pi}{1-\pi}\right) + \frac{\delta}{2\sigma
\sqrt{z^{\top}z}}\right],
%
\end{equation}
where $\Phi$ denotes the cumulative distribution function
of the standard normal. 
Similar steps can be carried out to obtain the probability
of correctly rejecting $H_0$, resulting in 
a very similar expression: 
\begin{equation}\nonumber
P(\text{reject } H_0 | \text{ } H_a \text{ is true})=1 - \Phi\left[\frac{\sigma\sqrt{z^{\top}z}}{\delta}
\text{ln}\left(\frac{K\pi}{1-\pi}\right) - \frac{\delta}{2\sigma
\sqrt{z^{\top}z}}\right]
\end{equation}
We now have all necessary components
to obtain a full expression of the goal function based on
the pre-specified cutoff, $P(H_0 | y ) \geq \frac{1}{1+K}$. 
Referring to Equation~\eqref{eq:rateofclass}, 
we can substitute all
fixed assignments and cutoff criteria to obtain 
\begin{multline}
\label{eq:goal_func}
G_B(n, \textbf{v}) = K\pi \Phi\left[\frac{\sigma\sqrt{z^{\top}z}}
{\delta}\text{ln}\left(\frac{K\pi}{1-\pi}\right) + \frac{\delta}{2\sigma
\sqrt{z^{\top}z}}\right]  + \\(1-\pi)
\left(1 - \Phi\left[\frac{\sigma\sqrt{z^{\top}z}}{\delta}
\text{ln}\left(\frac{K\pi}{1-\pi}\right) - \frac{\delta}{2\sigma
\sqrt{z^{\top}z}}\right]\right).
\end{multline}

\subsection{Goal Function in R Package}
The \fct{bayes\_goal\_func} function determines
the rate of correct classification in the context
of linear models discussed above. 
Specifically, the function determines the rate of correctly 
classifying the linear hypothesis expressions
as true or false. Recall the linear hypothesis test takes the form 
$H_0: u^{\top}\beta = c_0$ vs.  $H_a: u^{\top}\beta = c_1$.
The following set of parameters are to be specified 
prior to running the function: 
\begin{enumerate}
\item{\code{n} sample size (vector or scalar)}
\item{\code{Xn} design matrix that characterizes the data. This is
specifically given by the normal linear regression model 
$y = X\beta + \epsilon$,
$\epsilon ~ N(0, \sigma^2 I_n)$, 
where $I_n$ is an $n \times n$ identity matrix.
When set to \code{NULL}, an appropriate \code{Xn} is 
generated using \fct{bayesassurance::gen\_Xn}.}
\item{\code{K} the amount of utility associated with $H_0$ 
being correctly accepted. In this setting, $H_0$ is not rejected 
if the posterior probability of $H_0$ is at least
$\frac{1}{1+K}$.}
\item{\code{pi} constant corresponding to the prior probability
of observing $H_0: u^{\top}\beta = c_0$. Specifically,
$P(H_0) = 1 - P(H_a) = \pi$.}
\item{\code{u} fixed scalar or vector of the same dimension 
as $\beta_0$ and $\beta_1$}
\item{\code{sigsq} variance constant $\sigma^2$ contained in the 
linear regression model}
\item{\code{beta_0} fixed scalar or vector provided under
the assumption that $H_0$ is true such that $u^{\top}\beta=
u^{\top}\beta_0 = c_0$. }
\item{\code{beta_1} fixed scalar or vector provided under
the assumption that $H_a$ is true such that $u^{\top}\beta=
u^{\top}\beta_1 = c_1$. }
\end{enumerate}
The results are based on the solution derived
in Equation~\eqref{eq:goal_func}. Note that $z$ is
not directly specified as an input since the function
determines this value by solving the 
linear system $u = X^{\top}z$. This is specifically achieved
through the implementation of the Singular Value Decomposition 
on $X^{\top}$, enabling us to isolate $z$ by taking
the inverse on both sides of the equation.  

The next few examples demonstrate how this function
is used in the scalar setting and the vector setting. 

\subsubsection{Example 12: Example from Reference Paper}
\begin{figure}[h]
  \centering
    \includegraphics[width=0.5\textwidth]{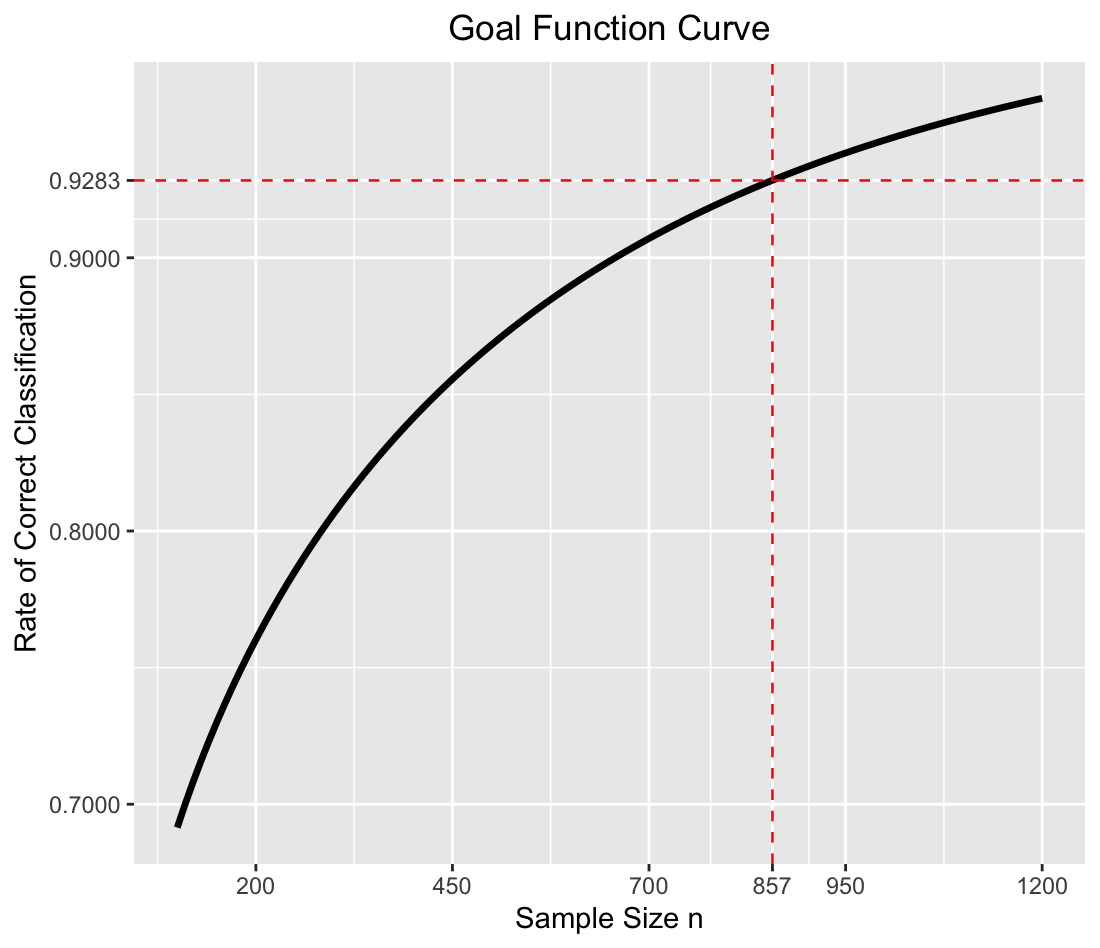}
    \caption{Rate of classification curve. Red dashed lines indicate a 
    sample size of n=857 is needed to ensure $r^{*}=0.9283$.}
  \label{fig:n_vs_r_curve}
\end{figure}

In this example, we implement the 
\fct{bayes\_goal\_func} function under a set of fixed
values taken from Section 2.1 of \cite{inoue}, 
where $\pi = 0.5$,  $K = 1$, $\sigma^2 = 1$
and the critical difference, $\delta$, 
is taken to be 0.1. We assign $\beta_0 = 0.5$ and $\beta_1 = 0.6$ 
to satisfy $\delta = 0.1$. 
The paper reports a minimum sample size of $n = 857$ 
is needed to ensure a rate of correct classification of 
$r^{*} = 0.9283$ under these settings.  Our results in 
Figure~\ref{fig:n_vs_r_curve}
support this claim by highlighting this exact set of
points on the curve.
The next block of code shows how to reproduce this figure, including
the red dashed lines using the \pkg{ggplot2} package. 
Notice that \code{u} is set to 1 since $\beta_0$ and $\beta_1$
are scalars. 

\begin{CodeChunk}
\begin{CodeInput}
R> library(bayesassurance)
R> library(ggplot2)

R> n <- seq(100, 1200, 10)

R> out <- bayes_goal_func(n, Xn = NULL, K = 1, pi = 0.5, u = 1, sigsq = 1, 
beta_0 = 0.5, beta_1 = 0.6)

R> head(out$rc_table)
R> out$rc_plot

## adding labels to ggplot
R> out$rc_plot + geom_hline(yintercept = 0.9283, linetype = "dashed", 
color = "red", size = 0.45) + geom_vline(xintercept = 857, 
linetype = "dashed", color = "red", size = 0.45) + 
scale_y_continuous(breaks = sort(c(seq(0.7, 1, 0.1), 0.9283))) + 
scale_x_continuous(breaks = sort(c(seq(200, 1250, 250), 857)))
\end{CodeInput}
\begin{CodeOutput}
    n Rate of Correct Classification
1 100                      0.6914625
2 110                      0.7000014
3 120                      0.7080588
4 130                      0.7156909
5 140                      0.7229434
6 150                      0.7298543
\end{CodeOutput}
\end{CodeChunk}

\begin{figure}[t!]
\centering
\includegraphics[width = 8 cm]{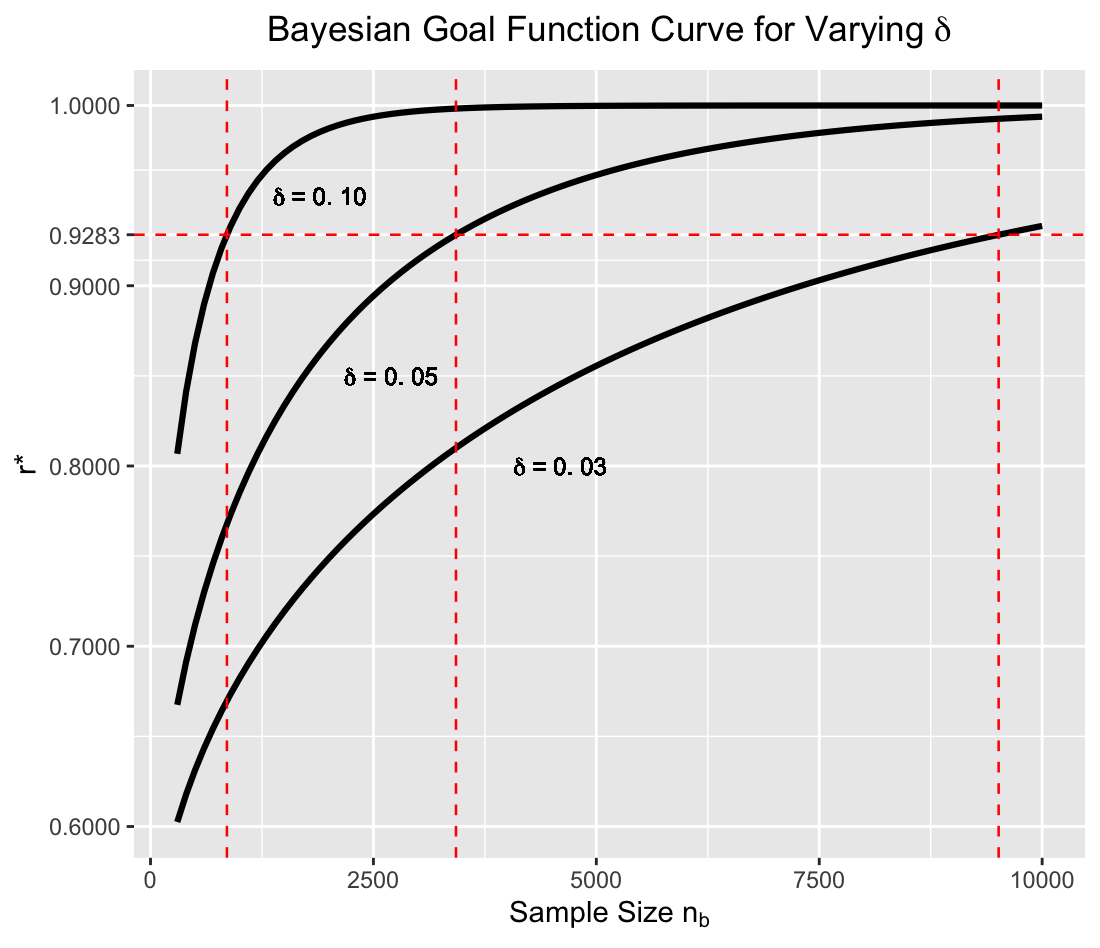}
\caption{\label{fig:ex12} Goal function curves resulting 
from the same set of fixed parameters and different
critical differences, $\delta$.}
\end{figure}

For further investigation,  we repeat the same steps
for two additional critical differences, displayed in 
Figure~\ref{fig:ex12}. The three intersections shown in the plot 
all correspond to the same $r^{*}$ threshold value, indicated
by the horizontal dashed line. We deduce that detecting
smaller critical differences requires larger 
sample sizes to meet the same rate
of correct classification criteria. Our results suggest that for critical
differences of $\delta = 0.10$, $\delta = 0.05$, and 
$\delta = 0.03$, the studies being implemented 
respectively need sample sizes of
857, 3426, and 9512 to satisfy $r^{*} = 0.9283$.  

\subsubsection{Example 13: Cost-effectiveness Application}
We revisit the cost-effectiveness application 
discussed in Section~\ref{subsec:ex2}.  Recall the net
monetary benefit formula, 
\[\xi = u^{\top}\beta = \Lambda(\mu_2 - \mu_1) - 
(\gamma_2 - \gamma_1),\]
which considers the efficacy, denoted by $\mu$, and costs,
denoted by $\gamma$, of the two treatments
undergoing assessment. The unit threshold
cost, originally denoted by $K$, is now represented as $\Lambda$
to limit any confusion from the constant $K$ that takes place in the
goal function. To re-express the net monetary benefit formula 
as $u^{\top}\beta$, we set $\beta = (\mu_1, \gamma_1, \mu_2, \gamma_2)^{\top}$ and
$u = (-\Lambda, 1, \Lambda, -1)^{\top}$. 

When analyzing this cost-effectiveness application
in previous examples, our primary focus involved determining 
the assurance of attaining $\xi  = u^{\top}\beta> 0$, which 
would indicate that Treatment 2 is more cost-effective 
than Treatment 1. The objective of the goal function, in contrast, 
estimates the probability of correctly identifying the true
outcome of a study. In this case, the underlying aim of our
goal function is characterized by correctly identifying
which treatment is more cost-effective than the other. 
We implement slight modifications to the original assurance
framework above that lets us assess
this from a goal function perspective. We 
separate the net monetary benefit formula 
into two components, $\xi_1$ and $\xi_2$:
\begin{align*}
\xi_1 = u^{\top}\beta_1 &= \Lambda\mu_1 - \gamma_1=
\begin{pmatrix}
\Lambda &
-1
\end{pmatrix}
\begin{pmatrix}
\mu_1 \\
\gamma_1
\end{pmatrix}\\
\xi_2 = u^{\top}\beta_2 &= \Lambda\mu_2 - \gamma_2=
\begin{pmatrix}
\Lambda &
-1
\end{pmatrix}
\begin{pmatrix}
\mu_2 \\
\gamma_2
\end{pmatrix}.
\end{align*}
It becomes clear how this modified framework
of the cost-effectiveness problem aligns with 
our goal function setting delineated above. Our hypothesis
test now takes on the linear form  $H_0: u^{\top}\beta = \xi_1$ vs. 
$H_a: u^{\top}\beta = \xi_2$.The user
needs to pass in $2 \times 1$ vectors
for \code{u},  \code{beta_0}, and \code{beta_1} to 
run this assessment. 
The following code segment demonstrates how this setting
is implemented in \proglang{R}. We use the same parameter
specifications from the example showcased in 
Section~\ref{subsec:ex2}. 

\begin{CodeChunk}
\begin{CodeInput}
R> library(bayesassurance)
R> library(ggplot2)

R> n <- seq(20, 100, 5)
R> Lambda <- 20000
R> sigsq <- 4.04^2
R> u <- as.matrix(c(Lambda, -1))
R> beta_0 <- as.matrix(c(5, 6000))
R> beta_1 <- as.matrix(c(6.5, 7200))

R> out <- bayes_goal_func(n, Xn = NULL, K = 1, pi = 0.5, u = u, sigsq = sigsq,
                       beta_0 = beta_0, beta_1 = beta_1)

R> head(out$rc_table)
R> out$rc_plot
\end{CodeInput}
\begin{CodeOutput}
     n Rate of Correct Classification
1   20                      0.7872786
2   25                      0.8135593
3   30                      0.8355023
4   35                      0.8541388
5   40                      0.8701601
6   45                      0.8840583
\end{CodeOutput}
\end{CodeChunk}

\subsection{Relationship to Frequentist Setting}
We can also construct an analogous goal function
using a classical-based approach by 
modifying the cutoff condition to one that adheres to 
measures pertaining to the frequentist setting. 
We continue working in a hypothesis 
testing framework within a general linear model setting 
outlined in Section~\ref{sec: goalfunc},
given as $H_0: u^{\top}\beta = u^{\top}\beta_0$ vs. 
$H_a: u^{\top}\beta = u^{\top}\beta_1$. 
As previously discussed, the Bayesian method 
formulates its goal function using the posterior 
probabilities of $H_0$ and $H_a$ to determine 
appropriate sample sizes $n_B$. 
In the frequentist setting, we rely on power to characterize the 
goal function.  To achieve a power of $1 - \beta$, the 
frequentist determines sample size $n_F$ using 
\begin{equation}
\label{eq:nf}
n_F = (z_{\alpha} + z_{\beta})^2\left(\frac{\sigma}{\delta}\right)^2, 
\end{equation}
where $\alpha$ and $\beta$ are respectively Type I and Type II
error rates and $\delta = u^{\top}\beta_1 - u^{\top}\beta_0$, 
denoting the critical difference.  

If we were not working within the context of general linear models
and were instead using sample means,
we could directly showcase the overlapping
behaviors exhibited between the Bayesian and 
frequentist settings simply by substituting $n = n_F$ 
into the goal function, $G_B(n, \textbf{v})$, given in 
Equation~\eqref{eq:rateofclass}. However, because our derived
expression of the rate of correct classification in 
Equation~\eqref{eq:goal_func} is not explicitly expressed 
in terms of $n$, we illustrate this relationship 
computationally through graphs.
Note that $n$ is included within the dimension
of the design matrix $X$, where $X$ is $n \times p$. 

We identify corresponding pairs of power and $r^{*}$ values
that yield equal sample sizes in the two settings. To determine
this using \proglang{R}, we code up a simple function 
for the frequentist sample size formula from 
Equation~\eqref{eq:nf}.
The function takes in a fixed set of parameters and returns the 
appropriate sample size that meets the specified conditions. 
Fixing the critical difference as $\delta = 0.1$, 
the significance level as 
$\alpha = 0.05$, and the variance as $\sigma^2 = 1$, 
we pass in a sequence of values for Type II error $\beta$ ranging between 0 and 1 to obtain the corresponding set of sample sizes,
which we save as \code{n_f} in the \proglang{R} script. We can 
subsequently determine the complementary set of 
$r^{*}$ values in the Bayesian
setting that are tied to the same set of sample sizes 
under the same fixed parameter assignments. The 
code block to perform these steps is documented below, 
along with the first ten rows of the data frame 
reporting  the exact sample sizes  being passed in and the 
specific pairs of $r^{*}$ and $\beta$ values that yield equal
sample sizes in the two settings. 

\begin{CodeChunk}
\begin{CodeInput}
R> nf_formula <- function(alpha, beta, sig_sq, delta){
  n_f <- ((qnorm(alpha) + qnorm(beta))^2)*(sig_sq/delta)^2
  return(n_f)
  }

R> beta_vals <- seq(0.01, 0.93, 0.01)
R> n_f <- sapply(beta_vals, nf_formula, alpha = 0.05, sig_sq = 1, delta = 0.1)
R> r_star <- bayes_goal_func(n = n_f, sigsq = 1, K = 1, pi = 0.5,
u = 1, beta_0 = 0.5,beta_1 = 0.6)
\end{CodeInput}
\begin{CodeOutput}
n = n_f = n_b  r_star    betas
1  1577.044136 0.9764596  0.01
2  1367.966073 0.9677457  0.02
3  1243.018843 0.9610338  0.03
4  1152.968984 0.9551570  0.04
5  1082.217382 0.9499830  0.05
6  1023.761436 0.9451140  0.06
7   973.842306 0.9405784  0.07
8   930.204365 0.9363449  0.08
9   891.385907 0.9322135  0.09
10  856.384735 0.9282491  0.10
\end{CodeOutput}
\end{CodeChunk}

\begin{figure}[t!]
\centering
\includegraphics[width = 9 cm]{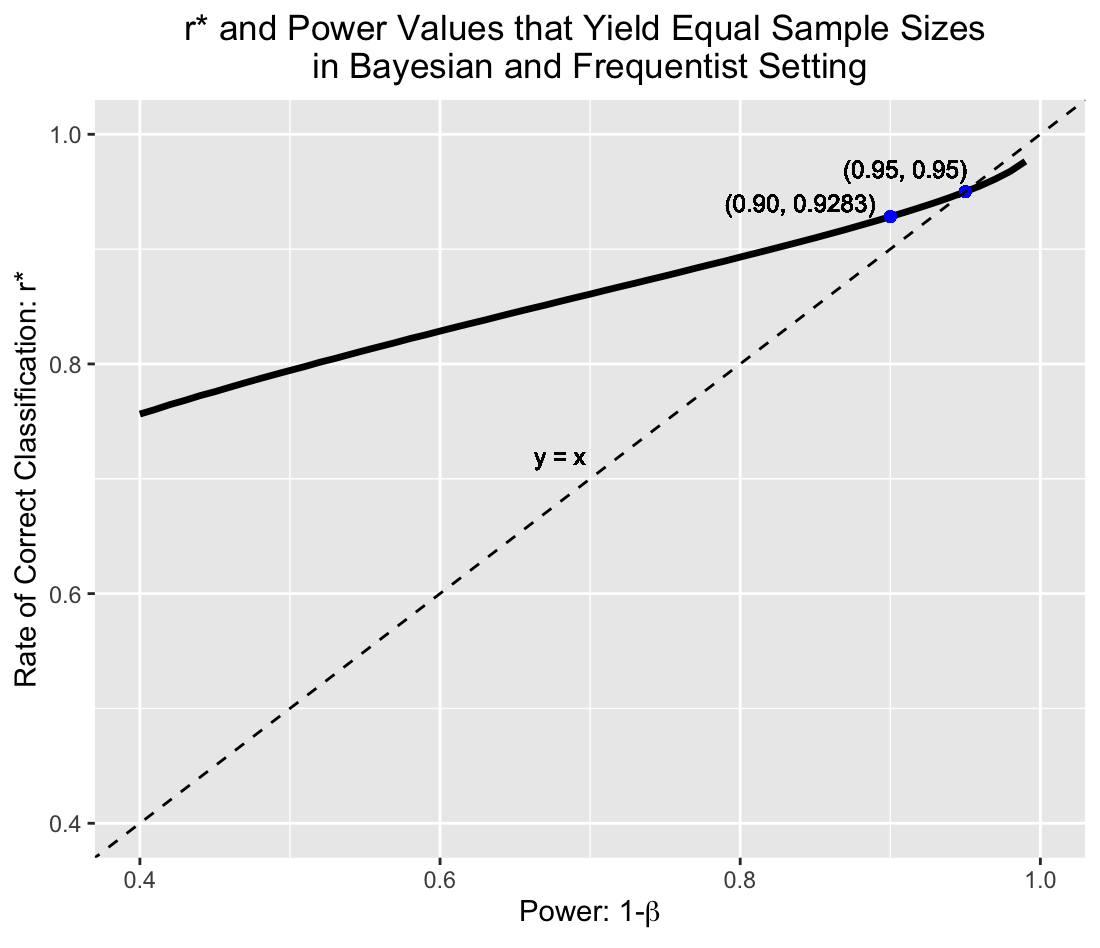}
\caption{\label{fig:ex13} Solid curve shows power and $r^{*}$
values that yield equal sample sizes in the Bayesian 
and frequentist settings assuming the critical difference
is $\delta = 0.1$. }
\end{figure}  

With the information gathered, we plot
the results in Figure~\ref{fig:ex13}, with power
on the x-axis and corresponding $r^{*}$ values on the y-axis.
Each point on the solid line indicates a $(1 - \beta, r^{*})$ 
point that yield equal sample sizes in the Bayesian and 
frequentist setting for the specified parameters
above. We include a reference line with slope = 1 and observe
larger discrepancies between the two values 
when power is small. The gap becomes visually smaller
as the power increases and the two lines coincide 
when both the frequentist power and rate of correct 
classification reach a value of 0.95, which corresponds to
a sample size of 1083.  This point is marked and labeled in the plot
along with another point we had previously referenced 
in Example 12. 

\section[Visualization Features and Useful Tools]{Visualization Features and Useful Tools} \label{sec:tools}

\subsection{Overlapping Power and Assurance Plots}

To facilitate the understanding of the relationship held between
Bayesian and frequentist settings, the \fct{pwr\_curves} function 
produces a single plot with the power curve and assurance points 
overlayed on top of one another.  Recall the 
primary difference held between
\fct{pwr\_freq} and \fct{assurance\_nd\_na}
is the need to specify additional precision parameters, 
$n_a$ and $n_d$, in \fct{assurance\_nd\_na}. 
Knowing that power and sample size analysis in the frequentist setting
is essentially a special case to the Bayesian assurance 
with precision parameters tailored to weak analysis priors
and strong design priors, the \fct{pwr\_curves} function
serves as a visualization tool in seeing how varying precision levels'
affect assurance values and how these assurance values 
compare to those of strictly weak analysis and 
strong design priors, i.e. power values.

\begin{figure}[t!]
\centering
\includegraphics[width = 10 cm]{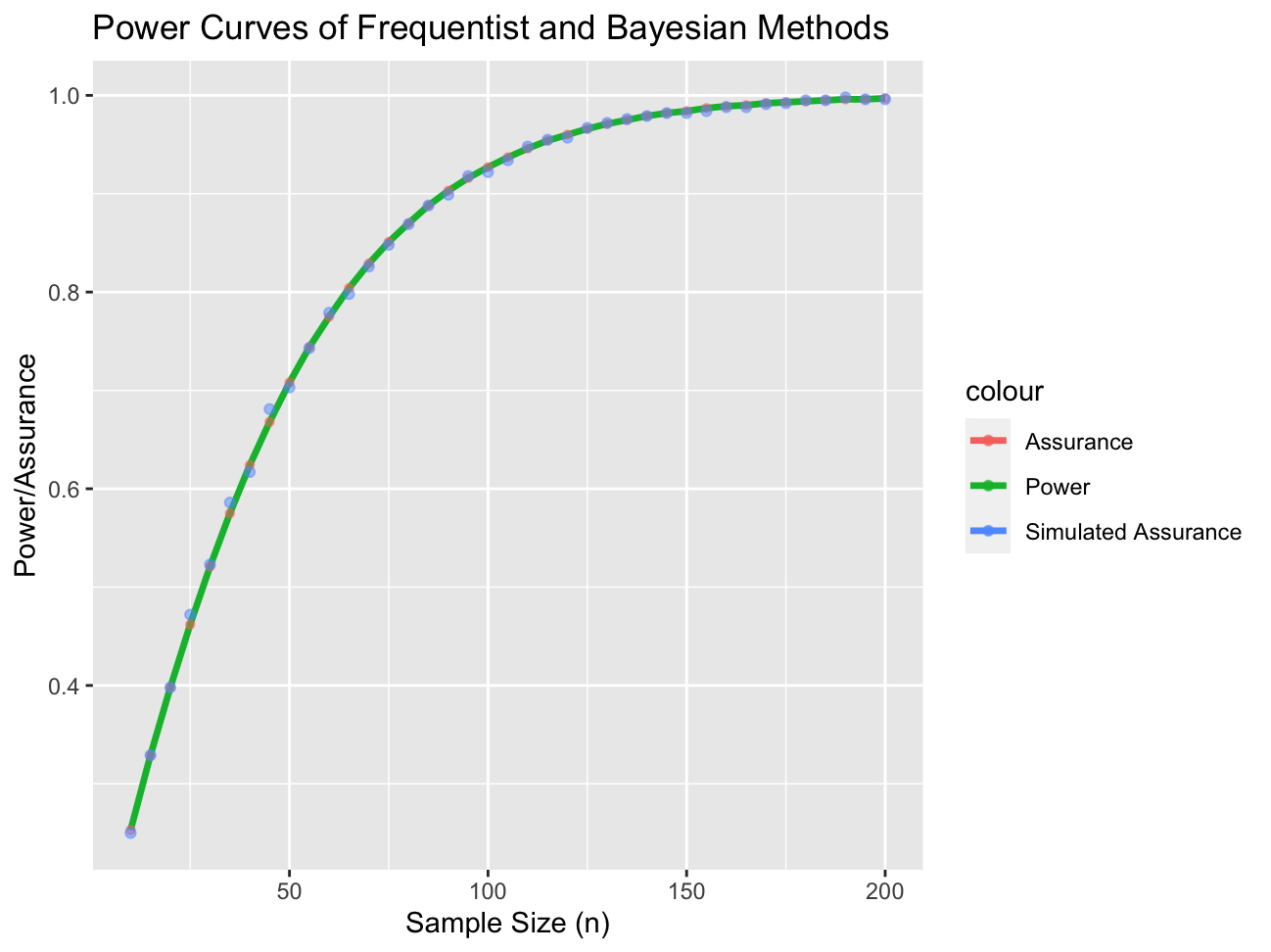}
\caption{\label{fig:pwr_curve_ex1} Power curve with exact and simulated
assurance points for weak analysis prior and strong design prior.}
\end{figure}

The \fct{pwr\_curves} function takes the combined set of
parameters presented in \fct{pwr\_freq} and \fct{assurance\_nd\_na},
which includes \code{n, n_a, n_d, theta_0, theta_1, sigsq,} and
\code{alpha}. For further customization, users have the option to
include a third set of points in their plot along with the power and
assurance curves. These additional points would correspond to the
simulated assurance results obtained using \fct{bayes\_sim}.
Optional parameters to implement this include 
\begin{enumerate}
	\item{\code{bayes_sim}: logical that indicates whether the 
	user wishes to include simulated assurance results obtained
	from \fct{bayes\_sim}.  Default setting is
	 \code{FALSE}.}
	\item{\code{mc_iter}: specifies the number of MC samples 
	to evaluate given \code{bayes\_sim = TRUE}.}
\end{enumerate}
The following code segment runs the \fct{pwr\_curves} function 
using a weak analysis stage prior (\code{n\_a} is
set to be small) and a strong design stage prior (\code{n\_d} is
set to be large). Implementing this produces a plot where the
assurance points lay perfectly on top of the power curve as
shown in Figure~\ref{fig:pwr_curve_ex1}. The simulated assurance
points obtained from \fct{bayes\_sim} are also plotted as we set
\code{bayes\_sim = TRUE}. These points are highlighted in blue, which 
lie very close in proximity to those of the exact
assurance points highlighted in red. 
We can also view individual tables of the three sets of 
points by directly calling them from the saved outputs, e.g.
\code{out\$power_table} shows the individual frequentist
power values for each sample size. The output we provide
shows the first ten rows. 

\begin{CodeChunk}
\begin{CodeInput}
R> library(bayesassurance)

R> out <- pwr_curve(n = seq(10, 200, 10), n_a = 1e-8, n_d = 1e+8, 
sigsq = 0.104, theta_0 = 0.15,theta_1 = 0.25, alt = "greater", alpha = 0.05, 
bayes_sim = TRUE, mc_iter = 5000)

R> head(out$power_table)
R> head(out$assurance_table)
R> out$plot

\end{CodeInput}
\begin{CodeOutput}
     n     Power
1   10 0.2532578
2   20 0.3981637
3   30 0.5213579
4   40 0.6241155
5   50 0.7080824
6   60 0.7754956

     n Assurance
1   10 0.2532578
2   20 0.3981637
3   30 0.5213579
4   40 0.6241155
5   50 0.7080824
6   60 0.7754956
\end{CodeOutput}
\end{CodeChunk}

\begin{figure}[t!]
\centering
\includegraphics[width = 8 cm]{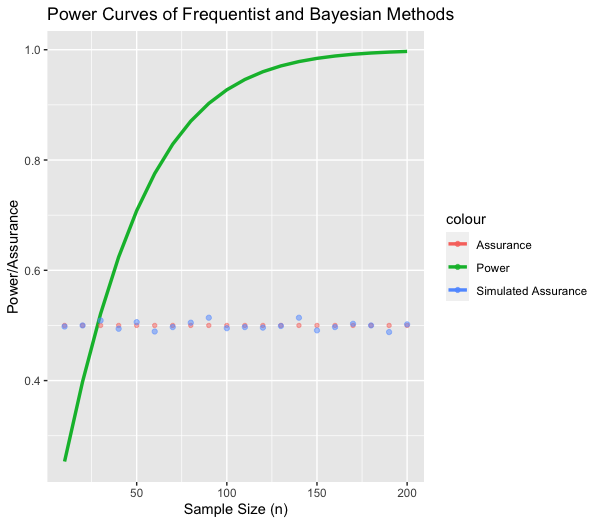}
\caption{\label{fig:pwr_curve_ex2} Power curve with exact and simulated
assurance points for weak analysis and design priors.}
\end{figure}

The next code segment considers the scenario in which both analysis
and design stage priors are weak (\code{n\_a} 
and \code{n\_d} are set to be small). This special case
shows how the assurance behaves when vague
priors are assigned. 
Substituting 0 in for both $n_a$ and $n_d$ in
Equation~\eqref{eq:assurance} results in a constant
assurance of $\Phi(0) = 0.5$ regardless of the sample size
and critical difference.  Figure~\ref{fig:pwr_curve_ex2} illustrates
these results, where we have the regular power curve and
the flat set of assurance points at 0.5 for both exact and simulated
cases. Note that some of these points appear purple
due to the overlaps that occur between the exact and simulated
assurance values. 

\begin{CodeChunk}
\begin{CodeInput}
R> library(bayesassurance)

R> pwr_curve(n = seq(10, 200, 10), n_a = 1e-8, n_d = 1e-8, 
sigsq = 0.104, theta_0 = 0.15,theta_1 = 0.25, alt = "greater", alpha = 0.05, 
bayes_sim = TRUE, mc_iter = 5000)

\end{CodeInput}
\end{CodeChunk}

\subsection{Design Matrix Generators}\label{design_matrix_generators}
The next sections go over design matrix generators
that run in the background of selected functions within
the \pkg{bayesassurance} package when the \code{Xn}
parameter is set to \code{NULL}.
We include these functions in case users wish to see 
how design matrices are constructed under this particular
setting. 

\subsubsection{Standard Design Matrix Generator}\label{subsubsec:standard_designmatrix}
The standard design matrix generator, \fct{gen\_Xn}, 
is relevant to a majority of the simulation-based assurance
functions discussed throughout the paper. 
It is first mentioned in Section~\ref{sec:simform} that
the assurance function under known variance,\fct{bayes\_sim},  
does not require users to specify their own
design matrix $X_n$. Users have the option of setting 
\code{Xn = NULL}, which prompts the function to construct a 
default design matrix using \fct{gen\_Xn} that complies
with the general linear model $y_n = X_n\beta +
\epsilon$, $\epsilon \sim N(0, \sigma^2V_n)$. The function
is automatically administered in the background while 
\fct{bayes\_sim} is actively in use.

When directly executed in an \proglang{R} script or console, 
the \fct{gen\_Xn} function takes in a single parameter, 
\code{n}, which can either be a scalar or vector. 
The length of \code{n} corresponds
to the number of groups being assessed in the study design
as well as the column dimension of the design matrix,
denoted as \code{p}. Therefore, in general, the resulting 
design matrix is of dimension $n \times p$. 
If a scalar value is specified for \code{n}, 
the resulting design matrix carries a dimension
of $n \times 1$. 

In the following example, we pass in a vector
of length $p = 4$, which outputs a design matrix of 
column dimension 4. Each column is comprised 
of ones vectors with lengths 
that align with the sample sizes passed in for
\code{n}. The row dimension is therefore the sum 
of all the entries in \code{n}. 
In this case, since the values 1, 3, 5, and 8 are
being passed in to \code{n}, the design matrix
to be constructed carries a row dimension of 
$1 + 3 + 5 + 8 = 17$ and a column dimension of 4. 

\begin{CodeChunk}
\begin{CodeInput}
R> n <- c(1,3,5,8)
R> gen_Xn(n = n)
\end{CodeInput}
\begin{CodeOutput}
      [,1] [,2] [,3] [,4]
 [1,]    1    0    0    0
 [2,]    0    1    0    0
 [3,]    0    1    0    0
 [4,]    0    1    0    0
 [5,]    0    0    1    0
 [6,]    0    0    1    0
 [7,]    0    0    1    0
 [8,]    0    0    1    0
 [9,]    0    0    1    0
[10,]    0    0    0    1
[11,]    0    0    0    1
[12,]    0    0    0    1
[13,]    0    0    0    1
[14,]    0    0    0    1
[15,]    0    0    0    1
[16,]    0    0    0    1
[17,]    0    0    0    1
\end{CodeOutput}
\end{CodeChunk}

The \fct{bayes\_sim} function and its related family of 
functions generate design matrices using \fct{gen\_Xn} 
in a very particular way. 
Each unique value contained in \code{n} that is
passed into \fct{bayes\_sim} corresponds to a distinct
study design and thus requires a distinct
design matrix. The \fct{gen\_Xn} function interprets each 
$i^{th}$ component of \code{n} as a separate balanced study 
design comprised of $n_i$ participants within each of the $p$
groups, where \code{p} is a parameter specified
in \fct{bayes\_sim}. For example, if we let 
\code{Xn = NULL} and pass in 
\code{n <- 2}, \code{p <- 4} for \fct{bayes\_sim},  
\fct{gen\_Xn} will process the vector
\code{n <- c(2, 2, 2, 2)} in the background. Hence, we'd obtain an 
$8 \times 4$ matrix of the form
\[
X_n = 
\begin{bmatrix}
1 & 0 & 0 & 0 \\
1 & 0 & 0 & 0 \\
0 & 1 & 0 & 0 \\
0 & 1 & 0 & 0 \\
0 & 0 & 1 & 0 \\
0 & 0 & 1 & 0 \\
0 & 0 & 0 & 1 \\
0 & 0 & 0 & 1 \\
\end{bmatrix}.
\]



\subsubsection{Design Matrix Generator in Longitudinal Setting}\label{subsubsec:long_designmatrix}
Section~\ref{sec:assur_longitudinal} describes how the 
linear model is extended to incorporate time-based 
covariates within the context of a longitudinal setting. For this
special case, a separate function is used to generate
design matrices that are appropriate for this setting. 
The \fct{genXn\_longitudinal} constructs
its design matrices differently than \fct{gen\_Xn} 
and therefore requires a different set of parameter specifications.
When the \code{longitudinal} parameter is
set to \code{TRUE} in \fct{bayes\_sim}, 
the user is required to specify the following set
of parameters, which are directly passed into 
\fct{genXn\_longitudinal}: 

\begin{enumerate}
\item{\code{ids}: vector of unique subject ids, usually of length 2
for study design purposes}
\item{\code{from}: start time of repeated measures for
each subject}
\item{\code{to}: end time of repeated measures for
each subject}
\item{\code{num_repeated_measures}: desired length of the repeated measures sequence. Should be a non-negative number, will be rounded up if fractional.}
\item{\code{poly_degree}:  degree of polynomial in longitudinal model, set to 1 by default.}
\end{enumerate}

Referring back to the model that was constructed for the 
case involving two subjects,  we observe in 
Equation~\eqref{eq:long_model} that the
design matrix contains vectors of ones within the first half
of its column dimension and lists the timepoints 
for each subject in the second half. Constructing this
design matrix requires several components. The user needs to
specify subject IDs that are capable of uniquely
identifying each individual in the study.  
Next, the user needs to specify the start and end time 
as well as the number of repeated measures reported 
for each subject. The number of repeated measures 
denotes the number of evenly-spaced timepoints that 
take place in between the start and end time. 
Since we are assuming a balanced longitudinal 
study design, each subject
considers the same set of timepoints. Finally, if the user
wishes to consider time covariates of higher degrees, such as 
a quadratic or cubic function, this can be altered using 
the \code{poly_degree} parameter, which takes on
a default assignment of 1. 

In the following code, we pass in a vector of 
subject IDs and specify the start and end
timepoints along with the desired length of the sequence.
The resulting design matrix contains vectors of
ones with lengths that correspond to the number of repeated
measures for each unique subject.

\begin{CodeChunk}
\begin{CodeInput}
 R> ids <- c(1,2,3,4)
 R> gen_Xn_longitudinal(ids, from = 1, to = 10, num_repeated_measures = 4)
\end{CodeInput}
\begin{CodeOutput}
      [,1] [,2] [,3] [,4] [,5] [,6] [,7] [,8]
 [1,]    1    0    0    0    1    0    0    0
 [2,]    1    0    0    0    4    0    0    0
 [3,]    1    0    0    0    7    0    0    0
 [4,]    1    0    0    0   10    0    0    0
 [5,]    0    1    0    0    0    1    0    0
 [6,]    0    1    0    0    0    4    0    0
 [7,]    0    1    0    0    0    7    0    0
 [8,]    0    1    0    0    0   10    0    0
 [9,]    0    0    1    0    0    0    1    0
[10,]    0    0    1    0    0    0    4    0
[11,]    0    0    1    0    0    0    7    0
[12,]    0    0    1    0    0    0   10    0
[13,]    0    0    0    1    0    0    0    1
[14,]    0    0    0    1    0    0    0    4
[15,]    0    0    0    1    0    0    0    7
[16,]    0    0    0    1    0    0    0   10
\end{CodeOutput}
\end{CodeChunk}

The next code block modifies the previous example 
to incorporate a quadratic term. Notice there are 
four additional columns being aggregated to the design matrix. 
These four columns are obtained from squaring
the four columns that precede this set of columns. 
\begin{CodeChunk}
\begin{CodeInput}
 R> ids <- c(1,2,3,4)
 R> gen_Xn_longitudinal(ids, from = 1, to = 10, num_repeated_measures = 4, 
 poly_degree = 2)
\end{CodeInput}
\begin{CodeOutput}
      1 2 3 4  1  2  3  4   1   2   3   4
 [1,] 1 0 0 0  1  0  0  0   1   0   0   0
 [2,] 1 0 0 0  4  0  0  0  16   0   0   0
 [3,] 1 0 0 0  7  0  0  0  49   0   0   0
 [4,] 1 0 0 0 10  0  0  0 100   0   0   0
 [5,] 0 1 0 0  0  1  0  0   0   1   0   0
 [6,] 0 1 0 0  0  4  0  0   0  16   0   0
 [7,] 0 1 0 0  0  7  0  0   0  49   0   0
 [8,] 0 1 0 0  0 10  0  0   0 100   0   0
 [9,] 0 0 1 0  0  0  1  0   0   0   1   0
[10,] 0 0 1 0  0  0  4  0   0   0  16   0
[11,] 0 0 1 0  0  0  7  0   0   0  49   0
[12,] 0 0 1 0  0  0 10  0   0   0 100   0
[13,] 0 0 0 1  0  0  0  1   0   0   0   1
[14,] 0 0 0 1  0  0  0  4   0   0   0  16
[15,] 0 0 0 1  0  0  0  7   0   0   0  49
[16,] 0 0 0 1  0  0  0 10   0   0   0 100
\end{CodeOutput}
\end{CodeChunk}

\section{Discussion}
This article introduced \pkg{bayesassurance},
a new \proglang{R} package that determines the Bayesian assurance
for various conditions using a two-stage framework.  
The goal of this package is to provide a convenient and user-friendly
implementation accessible to a wide range of data analysts, 
and showcase the generalized aspect of the Bayesian assurance
in relation to classical approaches to power and sample size 
analysis. We hope we have provided organized, well-documented
open-source code that can be used to address a wide selection
of clinical trial study designs and demonstrate the feasibility
of applying Bayesian methods to such problems. 

\bibliography{pow_samp}

\end{document}